# The Lunar Polar Icy Regolith and Possible Moonflows

Nick Gorkavyi [1], Oreste Reale [1], Harrison Schmitt [2]

[1] Science Systems and Application, Inc/GSFC/NASA, 10210 Greenbelt Road, Lanham, MD 20706
[2] P.O. Box 90730, Albuquerque, NM 87199-0730, USA

**Abstract.** The Lunar CRater Observation and Sensing Satellite (LCROSS) experiment, which created a crater 20–30 meters in size, showed that up to 6% water appear to be present in the permanently shadowed region (PSR) of Cabeus crater. Past independent theoretical analyses have concluded that some water molecules released elsewhere on the lunar surface can migrate to permanently shadowed areas. The estimated amount of ice, ice fragments, and other volatiles located in the polar regions in and beneath the regolith at depths of several meters may be still affected by significant uncertainties. Data from the Kaguya mission provide strong evidence for active plumes of water vapor erupting from polar regions outside the PSRs to heights of 3–18 km. These plumes appear to originate from subsurface ice located beneath the regolith surface at depths of approximately 10–20 cm. On the other hand, the polar highlands exhibit a relative deficit of craters larger than 25 meters, which may be caused by impact-melted ice lubricating regolith that flowed into initial crater depressions. At both the north and south poles of the Moon, signs of possible short-lived flows of liquid water-bearing material — moonflows — are presented in this work. Limited hydrous alteration of regolith ultra-fines to phyllosilicates may have occurred as these moonflows cooled.

## 1. Introduction

Many minerals in the Solar System contain water or hydroxyl (OH) groups within their crystal lattice. In chondritic meteorites that fall to Earth, the water content reaches several percent by mass. The surface layers of the Earth are relatively rich in water. In fact, even without considering the oceans, the research of Akhmanova, Dement'ev and Markov, (Akhmanova et al., 1979) demonstrated that even volcanic basalts contained about 0.5-1% crystallization water, which can be released from minerals under high temperatures within the Earth's mantle, and then raise upward together with other volatiles.

In contrast, from the late 19$^{th}$ century, scientists were convinced that the Moon is a very dry celestial body, and that any water on the surface would have been lost to space immediately because of low gravity and the lack of an atmosphere. Numerous sinuous channels found on the lunar surface were formed not by water flows but by volcanic lava flowing in lava tubes beneath hardened crust, with the path of the tubes gradually becoming exposed by impact rupture over time. There is no previously reported topographic evidence on the surface of the Moon that supports the idea of water-driven processes. The water content in the soil was estimated to be orders of magnitude lower than in terrestrial basalts. A remarkable dissentient opinion, developed by Watson, Murray and Brown, described in Watson et al., (1961), and later supported by Arnold (1979), argued on a theoretical basis that the amount of water lost along the entire lunar life must have been very small, and that the presence of ice in permanently shadowed regions (PSRs) of the Moon was possible. However, the dominant opinion of an anhydrous Moon kept persisting. The study of lunar samples returned by the Apollo missions did not change this view, even when lunar rock samples — including the famous 'orange soil', a type of pyroclastic ash (Meyer, 2012) brought back by Apollo 17 — were heated to temperatures of 150°–400°C and higher. The release of water vapor and carbon dioxide that was recorded (Cadenhead & Buergel, 1974a; Cadenhead & Buergel, 1974b) was attributed to possible contamination by terrestrial water, consequent to the fact that the containers used to return the samples were found to be non-hermetic because of exposure of seals to lunar dust. Later, Hauri et al (2011) documented the presence of water (0.06-0.14%) trapped in inclusions of olivine phenocrysts contained within glass beads in the Apollo 17 orange ash. Schmitt (2017, 2026) concludes that this pyroclastic water is derived from primordial sources in a chondritic lower mantle (below ~550 km).

The model of a dry Moon was supported by the dominant giant impact theory (e.g., Hartmann and Davis, 1975, Wood et al, 1986) caused by the collision of a Mars-sized object named Theia with the proto-Earth. According to this theory, the Moon was assumed to be a sphere of molten magma during its formation, from which most volatiles had escaped. During that phase, water was assumed to have been almost completely lost. In some version of the giant impact model, the Moon condensed from a nearly fully vaporized disk (Nakajima and Stevenson, 2014) of the Earth and Theia.

Alfvén and Arrhenius (1972) and the Safronov–Ruskol group (Safronov, 1972; Ruskol, 1975) developed the hypothesis of the accretional formation of the Moon, which did not assume such intense melting of the Moon at birth (see also Gorkavyi, 2023). This made the determination of the amount of water in lunar samples particularly important. Schmitt's indication of a chondritic lower mantle supports a lunar origin by accretion with only the upper ~550 km being a magma ocean produce by accretionary melting as potential and kinetic energy converted to heat.

Lunar core **s**amples (73001) vacuum sealed *in situ* and returned by the Apollo missions, were re-examined fifty years later (Shearer et al., 2024). In the study by Lucey et al. (2024), Apollo 17 samples were analysed and found to contain virtually no water, with a measured abundance of 0.005 ± 0.005%. The study also noted that lunar soil is highly hygroscopic and rapidly absorbs water from Earth's atmosphere, making contamination control a significant concern during sample handling and analysis. In addition, the process called 'lunar gardening', through which the upper lunar soil is stirred and mixed by micrometeorite impact, leading to a complete overturning across long temporal scales (Speyerer et al., 2016) is a likely mechanism to remove water.

Schmitt (2026) has demonstrated that the upper 3m of lunar regolith at Taurus-Littrow and probably elsewhere is made up of superposed regolith ejecta zones, generally 10-50 cm thick, deposited by identifiable source impact craters. As each regolith ejecta zone has been gardened over 20-300 Myr, it also is unlikely that there would be any significant residual water in nonpolar regions. It is important to emphasize that studies of samples collected from equatorial regions of the Moon do not provide direct constraints on the abundance of water in the polar regions, although the dynamics of regolith formation and gardening are likely the same or very similar. Under vacuum and high temperatures, a substantial portion of this surface water would have evaporated from the equatorial regions of the Moon. However, at the poles, under low-temperature conditions, especially in PSRs, this water in the form of subsurface ice may have been preserved, particularly if it were rapidly buried beneath a thermally insulating layer of regolith. According to calculations, in the polar regions, even in sunlit areas, ice can remain stable under the regolith at a depth of 1–2 meters (Gläser et al., 2021).

## 2. Missions that changed the understanding of Lunar Water

The year 1994 marks a turning point in lunar exploration, with resumed robotic missions by several countries. Particularly noteworthy was the mission Clementine, launched on January 25$^{th}$, 1994, as a joint project between the US Balistic Missile Defense Organization and NASA. Aside from being the first mission which provided a complete coverage of the Lunar surface, including the poles, it collected evidence of ice at the bottom of permanently shadowed craters at the Moon's south pole. Nozette et al. (1996), Spudis et al. (1998), Nozette et al. (2001), and McConnocchie et al. (2002), provided increasingly strong evidence of ice presence. In the wake of the interest triggered by Clementine, several other missions further expanded the confidence on the presence of lunar water and in the theoretical work of Watson et al. (1961) and Arnold (1979).

Partially conceived as a successor of Clementine, but also including instruments to measure gravity and magnetic fields, and evidence of radioisotopes, the Lunar Prospector mission **(**USA) was launched on January 7th, 1998, operated initially on a low polar orbit at about 100 km from the lunar surface, and was successively lowered to about 40 km, to be finally deliberately crashed in the crater Shoemaker on July 31st 1999. Relevant to this article, the mission conducted global mapping of the Moon using a neutron spectrometer (NS) to unveil presence of hydrogen up to depths of ~0.5 m. The most important result was the detection of enhanced hydrogen concentrations in the polar regions, particularly near Shackleton crater at the south pole, which was interpreted as evidence for water ice in PSRs even though it remained to be determined if solar wind hydrogen, present at a few hundreds of ppm in Apollo samples (Meyer et al, 2012) was even more concentrated at the poles. Estimates suggested that the upper regolith layer may contain up to ~1.5 wt.% water in the form of ice mixed with regolith (Feldman et al., 2001). Lunar Prospector data also indicated that the potential ice deposits near the north pole are comparable in magnitude to those at the south pole. Further analysis of neutron spectrometer measurements revealed regions with water contents of 0.03–0.08 wt.% at all latitudes (Lawrence et al., 2006; Lawrence et al., 2022).

A major leap forward in the understanding of Lunar volatiles was possible thanks to the Indian Space Research Organization (ISRO) Chandrayaan-1 mission, launched on October 22nd, 2008. (ISRO, 2008–2009) significantly changed the understanding of lunar volatiles. The NASA Moon Mineralogy Mapper ($M^3$) instrument, operating in the 0.43–3.0 μm spectral range, detected a broad absorption feature near 2.8–3.0 μm, interpreted as the presence of $OH/H_2O$ in the regolith across the lunar surface, with increasing abundance toward the poles (Pieters et al., 2009). Estimated water contents rise toward the polar regions to 0.05–0.075 wt.% (Li and Milliken, 2017). These observations provided the first strong evidence that smaller but significant accumulation of at least hydroxyl was not confined to PSRs, but widely spread on the surface of the Moon, although seemingly affected by the diurnal lunar cycle. The solar wind can serve as a source of hydrogen and hydroxyl in lunar regolith (Kling et al., 2025).

The 2009 Lunar CRater Observation and Sensing Satellite (LCROSS) NASA mission was specifically designed to further confirm the presence of water in permanently shadowed craters near the lunar south pole. Launched on 19 June 2009, the mission intentionally impacted the spent Atlas V Centaur upper stage (~2.2 t) into Cabeus crater (98 km diameter, 4 km depth; 84.9°S, 35.5°W) on 9 October 2009. The impact occurred at ~2.5 km $s^{-1}$, excavating roughly 350 t of lunar material and producing a crater ~27 m in diameter and ~5 m deep. The impact energy was equivalent to approximately 2 t of TNT. The collision generated a large ejecta plume reaching several kilometers in height. The LCROSS Shepherding Spacecraft, equipped with spectrometers and cameras, passed through the plume and analysed its composition before impacting the crater itself. Water was unambiguously detected in the ejecta plume in the form of $H_2O$ and OH. The measured water concentration in the crater was 5.6 ± 2.9 wt.%**,** increasing from 4.5 ± 1.4% shortly after impact to 7.1 ± 1.9% within the first 2–3 minutes (Colaprete et al., 2010). Strycker et al. (2013) estimated the water content in Cabeus Crater to be as 6.3 ± 1.6 wt.%. This experiment provided the first direct confirmation of water ice in lunar PSRs. This confirmed that the Moon, like Mercury, has polar ice accumulation.

The Lunar Reconnaissance Orbiter (NASA, operating since 2009) has produced the most detailed global dataset on the Moon. The Lunar Exploration Neutron Detector (LEND) onboard the LRO mapped hydrogen distribution via epithermal neutron measurements and confirmed enhanced hydrogen concentrations in permanently shadowed craters, although there are disagreements between the Lunar Prospector and LEND teams on data interpretations (Lawrence et al., 2006;

Lawrence et al., 2022). The Lyman Alpha Mapping Project (LAMP) instrument, also on the LRO, operating in the far-ultraviolet range (57–196 nm)**,** estimated water abundances of 0.9–4.9 wt.% in southern PSRs (Magaña et al., 2022). Consequent to these and other studies, interest in the topic of Moon water increased significantly (Crotts, 2011; Crotts, 2012a; Crotts, 2012b). Additional evidence of water being present on sunlit areas of the Moon was obtained by the Stratospheric Observatory for Infrared Astronomy (SOFIA), suggesting that the majority of water detected was contained in glassy minerals or voids between grains (Honnibal et al. 2021a).

Important information about lunar water was provided by the Japanese SELenogical and ENgineering Explorer (SELENE), also known as the Kaguya mission (Ohtake et al, 2024; Toyokawa et al, 2024). Spectra in the wavelength range of 513–1644 nm were analyzed in the twilight zone, where the lunar surface itself is dark while the near-lunar space remains illuminated (Ohtake et al., 2024). The analysis revealed localized and concentrated plumes consisting of mixtures of water and other vilatiles being ejected from the lunar surface into space in the polar regions. Absorption features associated with water and other volatile compounds were identified in 10,200 spectra obtained in non-permanently shadowed regions during lunar nighttime observations. Volatile signatures were repeatedly detected at certain locations, and most plume events showed no correlation with major meteoroid streams. Instead, enhanced activity was more frequently observed during the winter season at each pole. The timing, duration, geographic distribution, and thermal modelling of these events indicate that meteoroid impacts are unlikely to be their principal source. Rather, the observations suggest sublimation of volatile materials from subsurface regolith located at depths of approximately 10-20 cm. The interpretation that these water signatures originate from sublimation is also supported by the fact that the events are observed outside PSRs, since extensive sublimation is unlikely to occur within PSRs where temperatures remain continuously below 40K (Gläser et al. 2021). The observations indicate that the detected plumes rise to altitudes of at least 3 km and in some cases reach approximately 18 km above the lunar surface. In both polar regions, the number of detected volatile absorption events increased toward local midnight. Altogether, 72 plume events were identified during the 18-month observation campaign covering both poles.

The Chandrayaan-2 mission (2019–present) includes an orbiter operating in a ~100 km polar orbit. Its Imaging Infra-Red Spectrometer (IIRS)**,** covering 0.8–5.0 μm**,** has provided high-precision mapping of OH and $H_2O$ on the illuminated lunar surface at mid-latitudes (29°N–62°N), with water concentrations reaching ~0.008 wt.% (Chauhan et al., 2021). Sinha et al. (2026) identified evidence suggesting the presence of substantial subsurface ice deposits in doubly shadowed craters as revealed by Chandrayaan-2 dual frequency synthetic aperture radar.

These missions represent the major space experiments of recent decades that have significantly advanced our understanding of lunar water. The main results of lunar water research to date can be summarized in Table 1.

Table 1 Investigation of water content on the Moon

| Depths\Latitudes | Equator and mid-latitudes | Polar zones | Permanently shadowed craters |
|---|---|---|---|
| Surface (UV,IR) | ~0.01- 0.036% | 0.05 - 0.075% | 0.9 - 4.9% |
| <0.5 m (NS) | 0.03 - 0.08% | 0.1 - 0.15% | 1.5% |
| 0.35-0.7 m Apollo-17 (73001) | <0.01% | - | - |
| ~1.5 m (Akhmanova et al., 1979) | 0.1% | - | - |
| ~5 m (LCROSS) | - | - | 5.6% |
| Below ~500 m (primitive magma, Hauri et al, 2011; Schmitt, 2026) | 0.06-0.14% | - | - |

The relative abundance of water resulting from the above-mentioned missions is in apparent contradiction with the Moon impact theory as originally formulated. However, several attempts have been made to reconcile the impact theory with the concept of a 'wet Moon' such as Nakajima and Stevenson (2018) who argued about the inefficient loss of volatiles in the primordial disk supposedly originated by the impact or, among several others, Lock et al. (2018), who suggested that a phase of a super-rotating vaporized state (called synestia), containing Eart-originated vapor at elevated pressures, preceded the formation of the Moon. Aside from other impact-based theories which are consistent with the current Moon's composition and water presence, there are two additional possible models that align with experimental and observational facts:

1. The "Dry Moon with Snowdrifts" model**,** in which surface frost formed through the condensation of water vapor in shadowed craters. This water is external — it arrives on the Moon with comets and meteorites, evaporates upon impact, and part of the resulting water molecules settles into the "cold traps" of permanently shadowed craters, where temperatures are around −100°C and lower (Watson et al., 1961; Gläser et al. 2021). Additional sources of surface lunar water considered include protons associated with the solar wind, which can combine with OH, pyroclastic volatiles, and leakage from Earth's atmosphere (Kling et al., 2025; Kletetschka et al., 2022).

2. The "Moon with Underground Icy Regolith" concept. In this case, the Moon possesses primordial water (Watson et al., 1961) — like Earth, Mars, and possibly Mercury. If the Moon formed from material ejected from Earth's surface by powerful impacts, this does not necessarily mean it must be dry. Martian meteorites found on Earth were also ejected from Mars by strong impacts. As demonstrated by the discovery of significant amounts of water in the Martian meteorite NWA 7034 (Naver et al., 2026), such powerful impacts do not result in complete melting. Consequently, it is reasonable to expect that ejecta from Earth's surface into near-Earth orbit, produced by strong impacts, would not lead to complete evaporation of crystallization water, and the Moon could preserve part of the water delivered from Earth. Schmitt (1990, 2026), based on his geological studies, concluded that the Moon's interior was not incandescent but had moderate temperatures. During gravitational differentiation, primordial water rises from the hot lunar mantle into the cold crust but cannot persist at the surface due to evaporation or sublimation. However, at depths of several meters, water may remain in the form of ice. The existence of primordial water does not contradict the mechanism of frost accumulation in craters near the poles; on the contrary, it reinforces it: meteorite impacts on the lunar surface at the equator and mid-latitudes vaporize water contained in the soil, providing an additional contribution to the ballistic transport of water toward the polar regions.

In our view, the theory of lunar polar regions having subsurface ice as distributed particles in regolith may explain several observational facts:

1. The possibility that lunar water has an internal, primordial, origin is fully consistent with the results of the Moon Mineralogy Mapper ($M^3$), a NASA instrument onboard the Chandrayan-1 mission. It found evidence of magmatic water in the large and relatively young Bullialdus Crater (61 km in diameter, 3.5 km deep, 20.7°S, 22.2°W), located in Mare Nubium. $M^3$ determined a water content of 0.008% in the central peak of Bullialdus Crater (Klima et al., 2013). The impact that formed Bullialdus likely excavated material from depths of about 6–9 km into the lunar crust; therefore, one can conclude that a significant amount of lunar water is contained in the deeper layers of the Moon. Ground-based observations by the NASA InfraRed Telescope Facility (IRTF) confirmed the $M^3$ data, estimating a water content of 0.01–0.036% in the center of Bullialdus Crater (Honniball et al., 2021b). In addition, infrared observations by IRTF detected 0.005–0.015% water in the ejecta of the bright Aristarchus Crater (40 km in diameter, 2.7 km deep, 23.7°N, 47.4°W) in Oceanus Procellarum (Honniball et al., 2021b). Schmitt (2016, 2026) notes, however, that the overall geology of the Procellarum Basin may be complicated by mantle overturn and crustal plutons formed by resulting pressure-release partial melting and volatile concentration. The data on Bulliadus and Aritarchus Craters may need to be re-evaluated based on this interpretation. It also should be noted that there is an abundance of pyroclastic ash in the vicinity of Aristarchus (Wilhems, 1987).

2. The distribution of surface water and plume events correlates with the location of large permanently shadowed craters, but this relationship is not strict (Ohtake et al, 2024). For example, signs of water are visible in a mountainous region near coordinates −80.5° and 79°, as well as near Amundsen Crater, in areas around the coordinates −82.5° and 104°, and −84.5° and 135°.

3. In addition to water, the LCROSS experiment detected significant amounts of other volatiles in the ejecta plume produced above a crater 27 meters in diameter and 5 meters deep. The plume was intensely investigated by several authors in terms of dynamics and composition (e.g., Killen et al., 2010; Strycker et al., 2013). Colaprete et al., (2010) estimated that the relative abundances of other volatile molecules compared to water molecules were : hydrogen sulfide ($H_2S$) – 16.8%, ammonia ($NH_3$) – 6.0%, sulfur dioxide ($SO_2$) – 3.2%, ethylene ($C_2H_4$) – 3.1%, carbon dioxide ($CO_2$) – 2.2%, methanol ($CH_3OH$) – 1.6%, and methane ($CH_4$) – 0.7%. These gases have melting temperatures ranging from −75.5°C to −182.5°C — see Table 2:

Table 2. Melting and boiling (sublimation) temperatures of gases detected in Cabeus Crater.

| | $H_2O$ | $H_2S$ | $NH_3$ | $SO_2$ | $C_2H_4$ | $CO_2$ | $CH_3OH$ | $CH_4$ | $N_2$ |
|---|---|---|---|---|---|---|---|---|---|
| Abundances of volatile molecules compared to water molecules (%) | 100 | 16.8 | 6.0 | 3.2 | 3.1 | 2.2 | 1.6 | 0.7 | 0 |
| Melting temperature (°C) | 0 | −85.5 | −77.7 | −75.5 | −169.2 | - | −97.6 | −182.5 | −209.9 |
| Boiling/sublimation temperature (°C) | 100 | −59.6 | −33.3 | −10.0 | −103.7 | −78.5 | 64.7 | −161.5 | −195.8 |

The mass fraction of gases (excluding water) detected in the Cabeus crater ejecta is about 3.5% of the soil mass. Together with water, the total volatile content in the Cabeus regolith can be estimated ~ 9-10%. As shown by Watson, Murray, and Brown in their paper (Watson et al., 1961), over geological timescales only a few grams of water ice per square centimeter would sublimate from the surface of a permanently shadowed region (PSR) with a temperature of 120 K. The same study demonstrated that the evaporation rates of carbon dioxide, ammonia, and sulfur dioxide are many orders of magnitude higher. This implies that these volatiles could not have remained stable on the lunar surface over geological timescales and therefore were likely excavated from depths of several meters by the impact event.

The origin and reserves of lunar water will become clear after isotopic analysis of soil samples taken in the polar and other regions of the Moon from depths of 10–20 meters, which will require delivering a compact drilling rig on our satellite. It is quite possible that the unfilled cells of the Table 1 conceal important surprises. The possible existence of a significant ice component in polar regolith compels us to examine carefully the geomorphological structures of the lunar polar regions: are there signs of a subsurface layer of water and volatiles there?

## 3. Geomorphological Signs of Slow Ice Flow in the Lunar Polar Regions

The geomorphology of the lunar polar regions indicates active surface renewal processes (Figure 1). In the lunar South Polar region, heavily cratered dark plains frequently occur alongside relatively smooth, bright highlands (Figure 1). In Figure 1, one can clearly see how significantly the blurred forms of polar craters differ from the dry, sharply degraded craters typical of equatorial and mid-latitude regions. It should be noted that craters degrade and disappear much more rapidly on lunar slopes than on relatively flat surfaces due to gravitationally asymmetric ejecta distribution that accelerates downslope movement of regolith (Schmitt, 2026). How might the enhanced abundance of ice and other volatiles influence the evolution of polar topography?

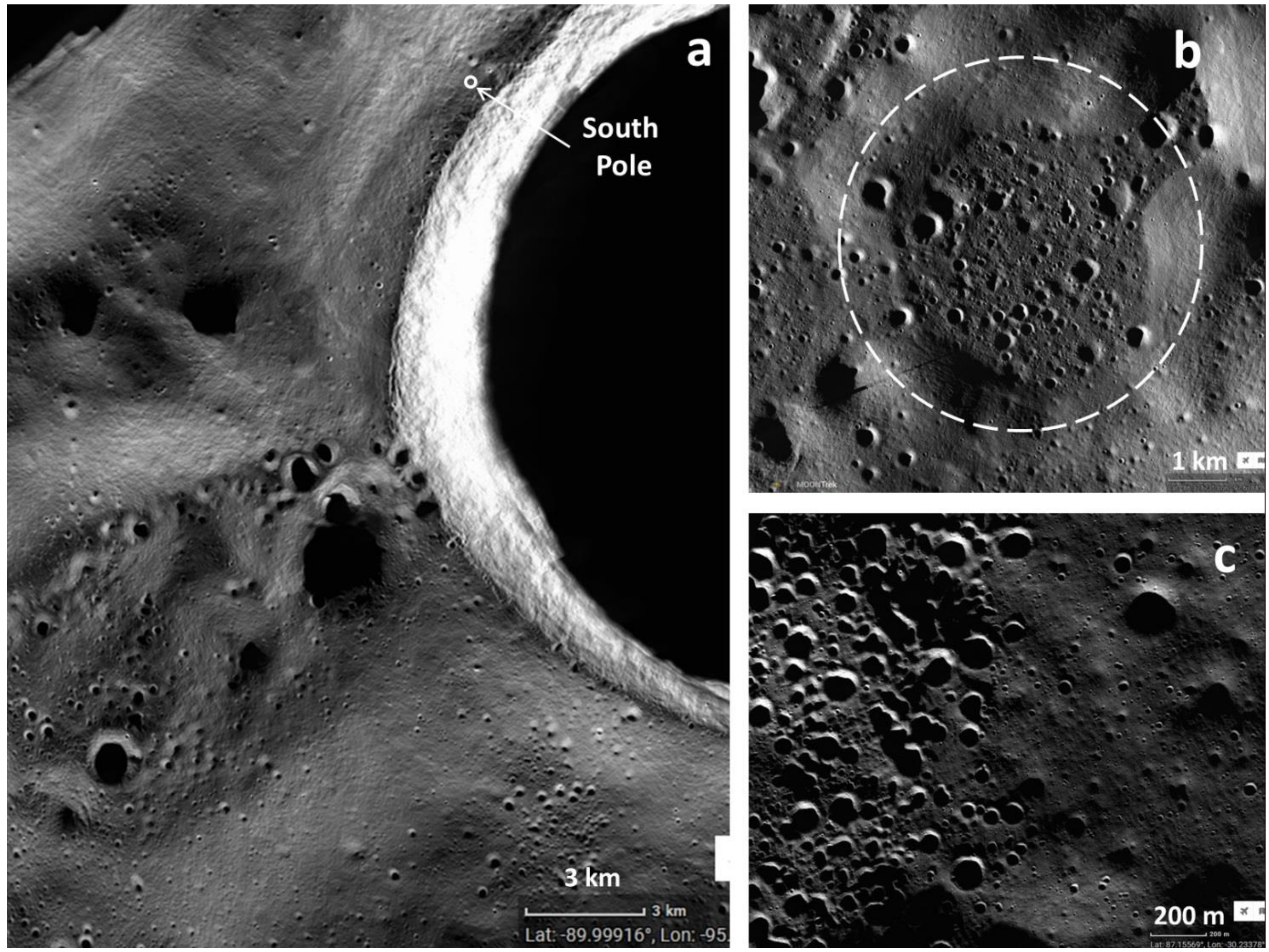

Figure 1 Craters of the Lunar South Pole. (a) The diversity of lunar terrain near Shackleton Crater, on whose rim the South Pole is located. (b) An unusual unnamed crater approximately 5–6 km in size (coordinates −86.36°S, −97.8°W, outlined with a dashed line) may be a highly degraded old impact crater that has been partially filled with lava or possibly a coincidental circular alignment of other hilly features. Inside the crater are numerous small craters (100–500 m), while along its edges lie elevated areas with a relatively young appearing surface. (c) Two areas exhibiting different crater densities near the lunar North Pole. The image covers a region measuring 2 × 1.6 km centered at 87.2°N, 30.2°W. NASA photo (MoonTrek).

On Earth, glaciers flow at speeds ranging from several meters to several kilometers per year, depending upon topography, environmental conditions, base water lubrication, and ice thickness. At low temperatures, the viscosity of ice (if not compensated by pressure-induced partial melting) increases exponentially with decreasing temperature, yet over billions of years, its plasticity should still be sufficient to allow movement over hundreds of meters (Sori et al., 2016). While the Moon's low gravity is expected to substantially reduce the rate of ice flow relative to Earth, the dominant control on ice rheology is the exponential temperature dependence of viscosity. Consequently, viscous flow can become an important geological process even on small bodies. For example, the flow of volatile ices (primarily $N_2$, CO, and $CH_4$) has profoundly shaped the surface of Pluto (Moore et al., 2016).

Therefore, in the lunar polar regions, craters in ice-dominant regolith, if present and particularly on slopes, could gradually disappear due to the flow of subsurface ice. An additional sign of ice motion could be a system of multidirectional fractures, (extensional graben and tensional shear fractures) analogous to the crevasse and fracture patterns observed in creeping terrestrial glaciers. In regolith in which ice is not dominant as an enclosing matrix, no such flow would be expected as ice would be just another mineral particle.

For polar lunar latitudes, dark zones with a high density of craters are typical, as well as brighter regions where craters are much less abundant (Figure 2).

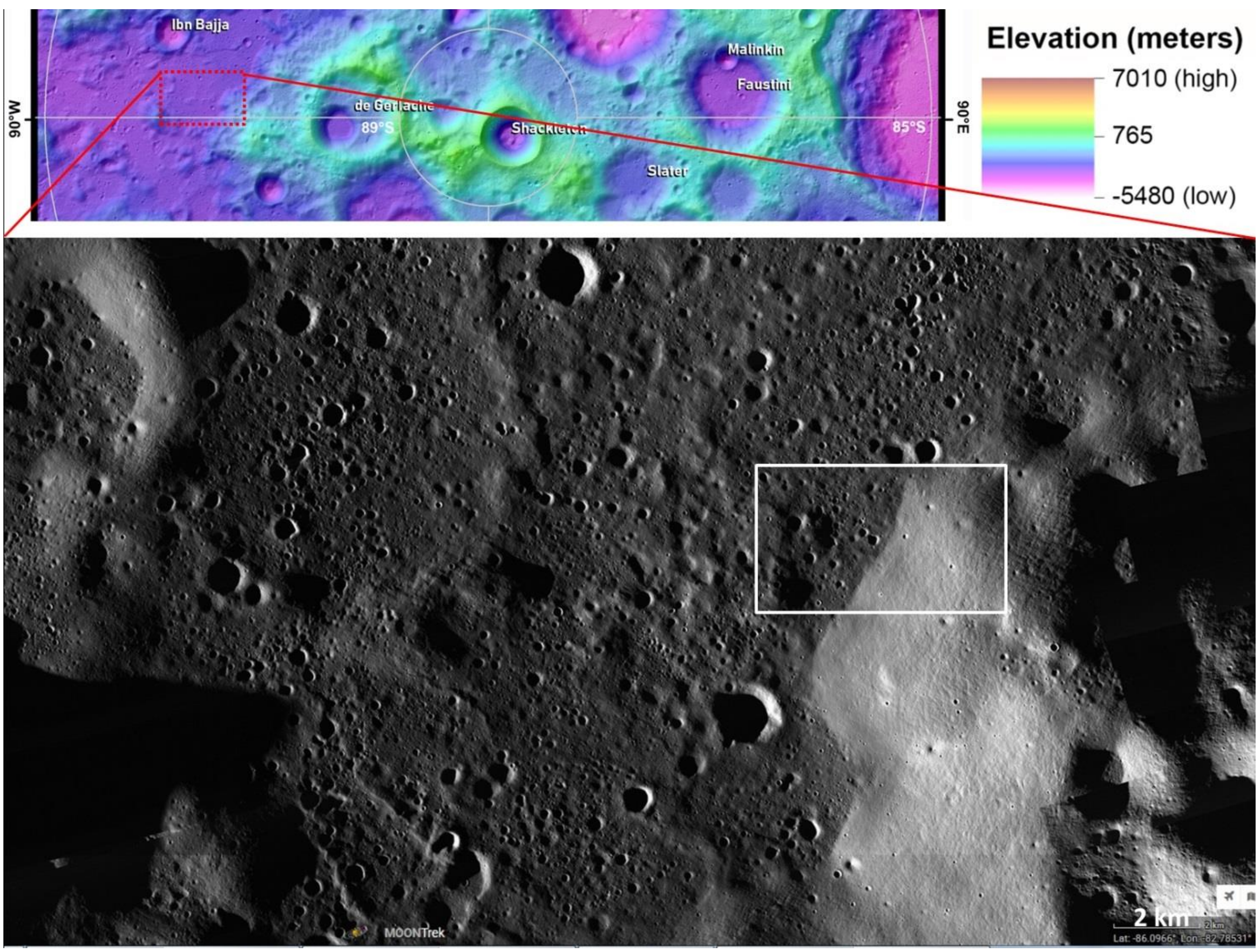

Figure 2 Top: Topographic map of the lunar South Polar region. Bottom: A 30 × 17 km area located at the boundary between the mare-like plain and the mountainous terrain (the lower-left corner is at 87.3°S, 91.3°W). NASA/Moon Trek. It is evident that the mountainous regions exhibit a much smoother surface, whereas the lower-lying plain is densely covered with impact craters. The white rectangle indicates the area examined in Figure 3.

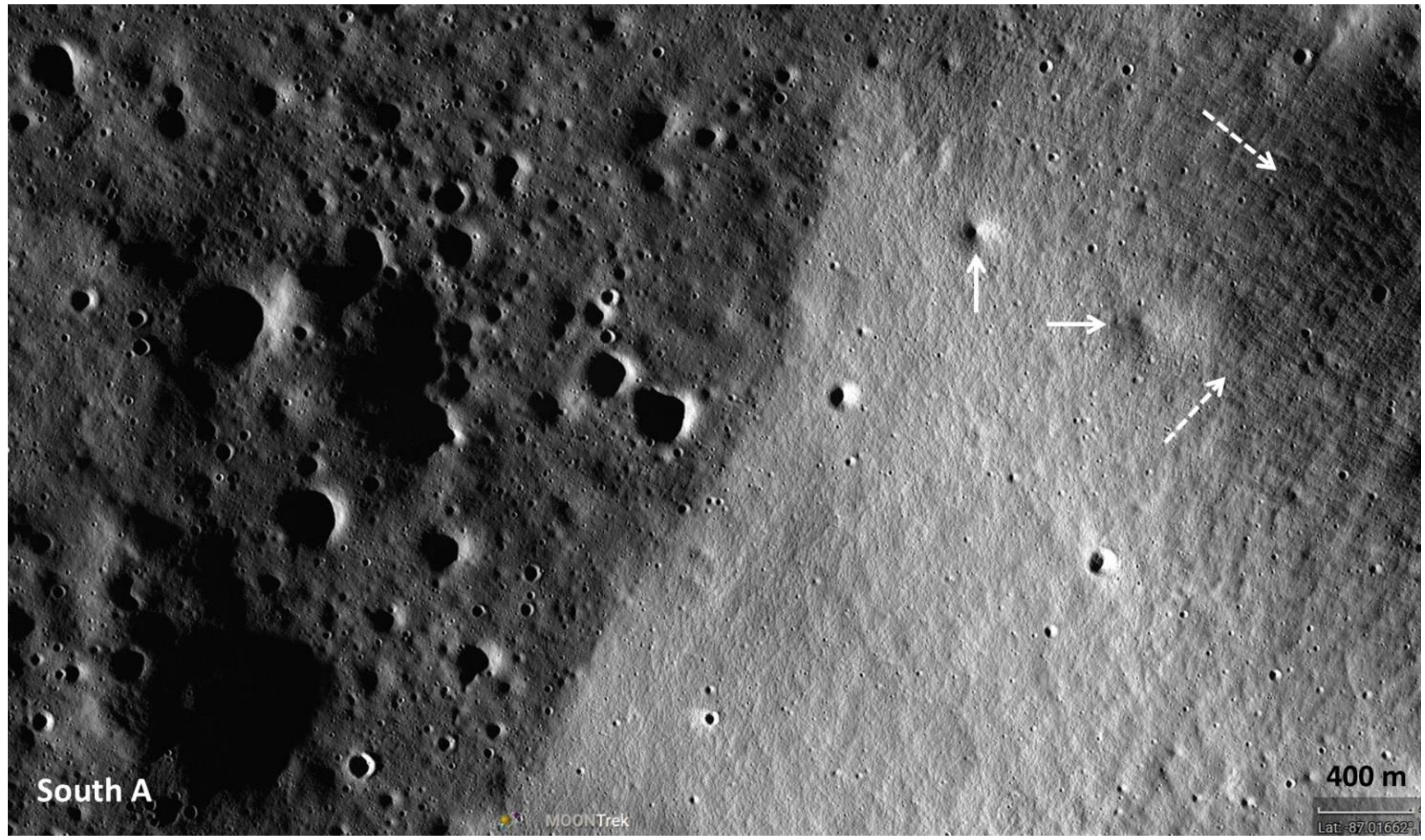

Figure 3. Region near the South Pole (-87.0°S, -86.4°W; 3.5 km x 6 km). Note sharp contrast between the dark, relatively level area on the left shown in Fig. 2, and light slope areas on the right that is similar to contacts between plains and slopes in lower latitudes where downslope accumulation of regolith produces a reduction in slope over a few hunred meters. Light area on the right has ridge and trough pattern similar to the light mantle avalanche at Taurus-Littrow (Schmitt 2026); however, its ridge and trough pattern may reflect the variations in the dynamics of maturation as a function of varying slopes rather than those of a flowing and settling avalanche. Rimless, pit craters with D ~ 100 m (white arrows) exhibit a rimless appearance, possibly indicating that the surrounding regolith has flowed inward into crater formed on a ridge line as suggested by the variation in photometric effects. These pit craters may be where fine avalanche material drained into old craters or their shape may reflect a thiner regolith cover along the ridgeline. The dashed arrows indicate an oblique ridge and trough texture with a crossing tensional separation related to downslope migration of regolith. These structures are also characteristic of permafrost regions on Earth (Gorkavyi, 2023). NASA (MoonTrek)

Figure 3 shows a region 21 km² where an old, heavily cratered area on the left and a younger area on the right. Each surface type occupies approximately half of the studied area. By a *young surface*, we mean a region where resurfacing processes are active (for example, crater relaxation or infilling), even though the underlying terrain itself may be very old. Let us assume that both regions shown in Figure 3 have the same age of first surface exposure of about 4 billion years. In order to fill a crater 400 meters in diameter over 4 billion years (see Figure 3), the ice would need to flow at a rate of approximately 0.05 microns per year. For comparison, in the polar regions of Mars at temperatures around 170 K, the relaxation rate of a 200-meter crater due to ice flow is about 0.1 microns per year for horizontal flow and up to 0.4 microns per year for vertical flow (Sori et al., 2016). Although Martian gravity is about twice that of the Moon, which accelerates ice flow, temperatures at depths of several meters in illuminated lunar regions are noticeably higher than those assumed for the Martian crater model. Overall, our estimate is reasonably close to the inferred flow rates of Martian icy regolith.

Table 3 presents the results of a visual count of craters of different sizes. We refer to the 21 km² polar region shown in Figure 3 as *South A***.** The table also includes data for two additional polar regions: *South B* (-86.06°S, -87.18°W) and *North A* (87.19°N, -74.55°W), each covering an area of 12.6 km².

Table 3 Crater statistics for the polar regions

| **Region / crater size** | **10-25 m** | **25-50 m** | **50-150 m** | Polar area |
|---|---|---|---|---|
| Bright highland | 346 | 40 | 19 | South A (10.5 sq.km) |
| Dark plain | 789 | 149 | 91 | South A (10.5 sq.km) |
| **Ratio** | **0.44** | **0.27** | **0.21** | South A |
| Bright highland | 211 | 30 | 12 | South B (6.3 sq.km) |
| Dark plain | 288 | 105 | 35 | South B (6.3 sq.km) |
| **Ratio** | **0.73** | **0.29** | **0.34** | South B |
| Bright highland | 167 | 28 | 8 | North A (6.3 sq.km) |
| Dark plain | 239 | 77 | 24 | North A (6.3 sq.km) |
| **Ratio** | **0.78** | **0.36** | **0.33** | North A |
| Bright highland | 744 | 98 | 39 | Sum (23.1 sq.km) |
| Dark plain | 1316 | 331 | 150 | Sum (23.1 sq.km |
| **Ratio** | **0.57** | **0.30** | **0.26** | Sum |

The number of distinct impact craters on the dark and bright halves of the 21 km² area are distinctly different; however, craters on the slopes in the light half degrade at a higher rate than in the dark half. Therefore, the difference in crater density per unit area between the two regions has been determined by crater lifetime, particularly as the light area is significantly older terrain than the dark area and should have a higher crater frequency than the latter. For lunar regions where craters are gradually filled by impact ejecta from other craters, a well-known rule applies: small craters are filled by regolith ejecta faster than larger ones. The mechanism behind this rule is straightforward: the volume of a crater bowl depends on the cube of the crater radius, whereas the area through which regolith enters the crater depends only on the square of the radius. Consequently, larger craters fill more slowly. The lifetime of craters 10–100 m in diameter should be proportional to their radius (or depth), which is confirmed by lunar crater counts and theoretical modeling that even suggests a nonlinear dependence of crater lifetime on size, approximately $\sim D^{(1.1-1.3)}$ (Fassett et al., 2022). These relationships apply only to regions with similar geomorphology, for example, plains. On mountain slopes, impact craters degrade several times faster than on flat terrain.

The Taurus–Littrow Valley (20°N, 31°E) also contains two contrasting terrain types: a dark, heavily cratered plain and relatively smooth, bright highlands represented by the North Massif (Figure 4). To compare the crater densities of the polar regions with those of the Taurus–Littrow Valley, we performed the same statistical analysis for two areas (Region C, 20 km², and Region D, 14 km²), in which the dark plains and bright highlands occupy approximately equal areas. The results of the analysis are presented in Table 4.

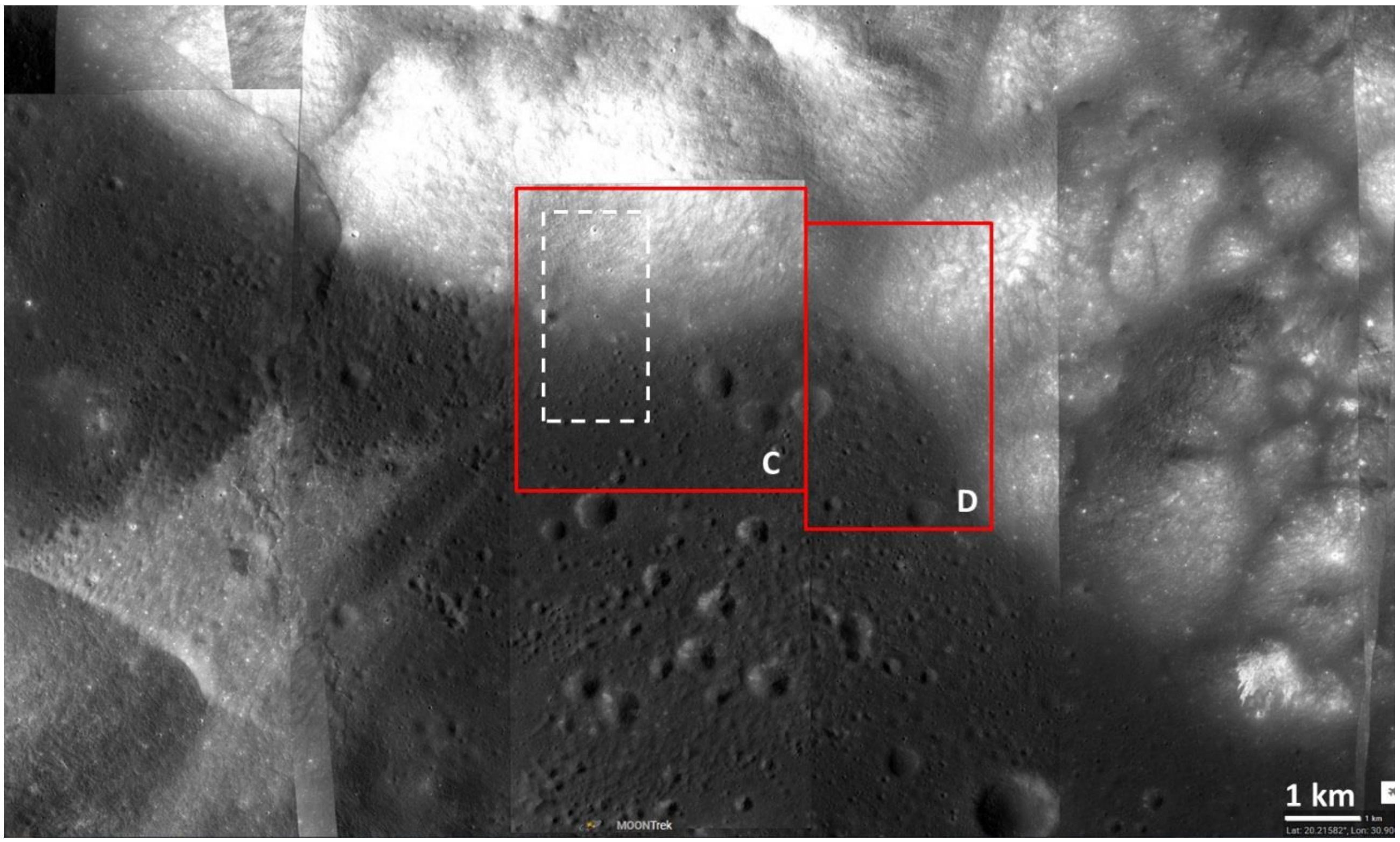

Figure 4. The Taurus–Littrow Valley and the North Massif. The two regions analyzed statistically are outlined in red. The hatched area corresponds to the region shown in Figure 5.

Table 4 Crater statistics for the Taurus-Littrow region

| **Region / crater size** | **10-25 m** | **25-50 m** | **50-150 m** | Taurus-Littrow |
|---|---|---|---|---|
| Bright highland | 317 | 26 | 5 | Zone C (10 sq.km) |
| Dark plain | 482 | 29 | 28 | Zone C (10 sq.km) |
| **Ratio** | **0.66** | **0.90** | **0.18** | Zone C |
| Bright highland | 463 | 46 | 10 | Zone D (7 sq.km) |
| Dark plain | 654 | 57 | 39 | Zone D (7 sq.km) |
| **Ratio** | **0.71** | **0.81** | **0.26** | Zone D |
| Bright highland | 780 | 72 | 15 | Sum (17 sq.km) |
| Dark plain | 1136 | 86 | 67 | Sum (17 sq.km) |
| **Ratio** | **0.69** | **0.84** | **0.22** | Sum |

The polar regions examined in this study are older than the Taurus–Littrow Valley. Evidence for this is provided by the fact that the density of craters 25–150 m in diameter on the dark polar plains is 2.3 times higher than on the dark lowland floor of the Taurus–Littrow Valley (Tables 3 and 4). In contrast, the density of craters of the same size on the bright polar highlands is only about 15% higher than that on the North Massif of the Taurus–Littrow Valley. Thus, craters on the polar highlands appear to have undergone substantially greater degradation than those on the North Massif. On the North Massif the number of craters 50–150 m in diameter is nearly five times lower than on the lowland floor of the Taurus–Littrow Valley, whereas the abundance of smaller craters is reduced by only 10–30% (Table 4). This suggests that the degradation processes operating on the slopes of the North Massif are highly effective for craters larger than about 50 m but much less effective for smaller craters.

The polar regions exhibit a different pattern. Craters with diameters between 25 and 50 m are depleted approximately 2.8 times more strongly than on the slopes of the North Massif (Table 3). In other words, both small (10–25 m) and large (50–150 m) craters on the polar highlands appear to degrade at rates comparable to those on the North Massif, whereas intermediate-sized craters (25–50 m) undergo substantially more rapid degradation. We cautiously suggest that this behavior may be related to the presence of an ice-rich regolith layer beneath the polar highlands at depths of several meters.

Figure 5 shows the two polar regions, South B and North A, together with the study area in the Taurus–Littrow Valley (outlined by a dashed line in Figure 4). The polar regions differ noticeably from the equatorial study area not only in the distribution and morphology of impact craters, but also in the remarkably uniform albedo of their highland surfaces.

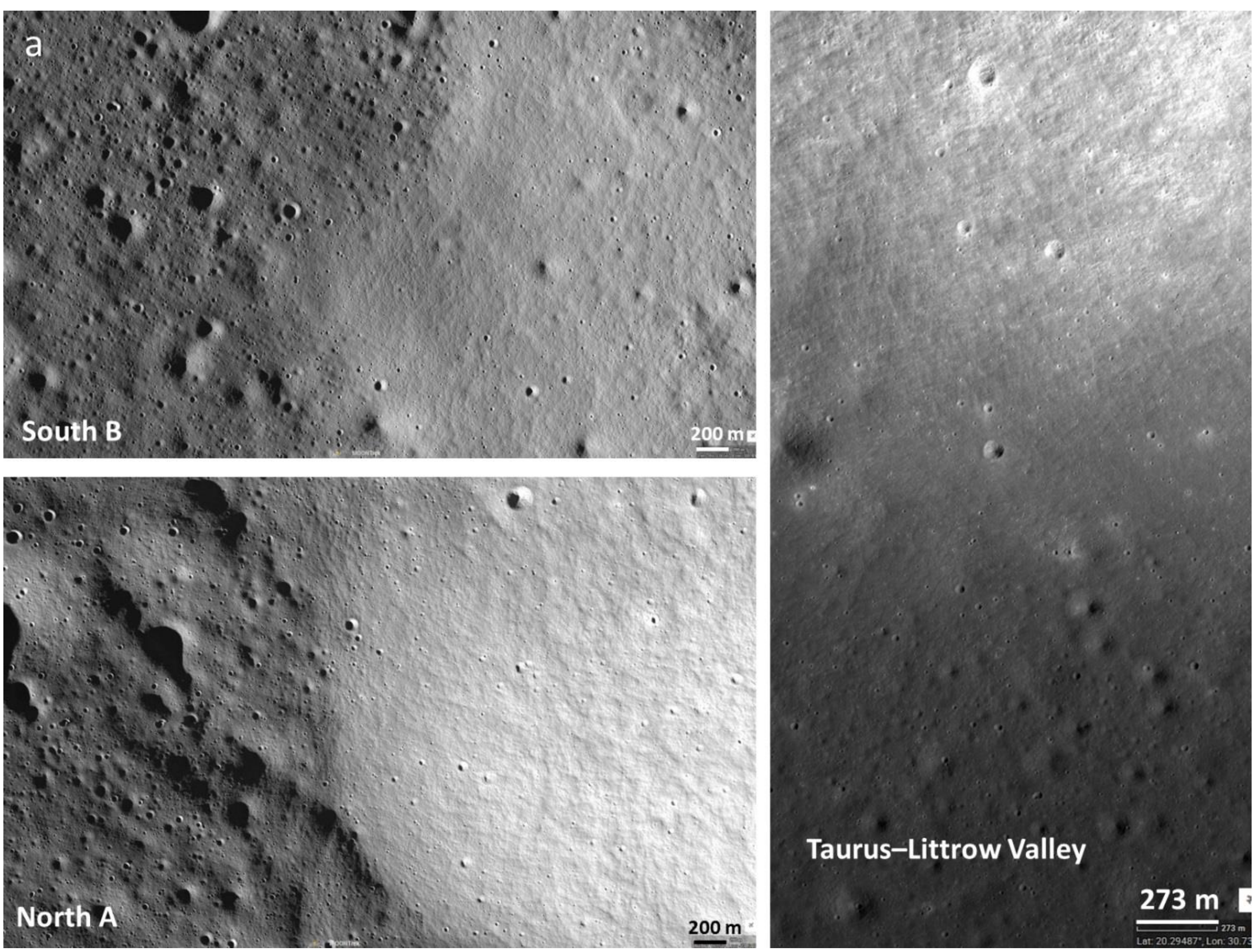

Figure 5. The figure shows the *South B* and *North A* polar regions (with adjusted brightness) used for the statistical analysis, together with the study area in the Taurus–Littrow Valley (outlined by a dashed line in Figure 4).

Geomorphological signs of enhanced subsurface volatile content in the lunar polar regions include not only the gradual infilling of small craters due to ice flow, but also numerous landslides and slope fractures analogous to crevasses observed in flowing terrestrial glaciers (Gorkavyi, 2022; Gorkavyi, 2023) – see Figure 6. Similar fractures are also found in equatorial regions of the Moon, because they can form not only from ice motion but also from regolith creep due to micro-meteor impacts, seismic disturbances, and thermal stress over hundreds of millons of years. Therefore, more definitive conclusions about the nature of fractures in the lunar polar regions will require detailed future investigations.

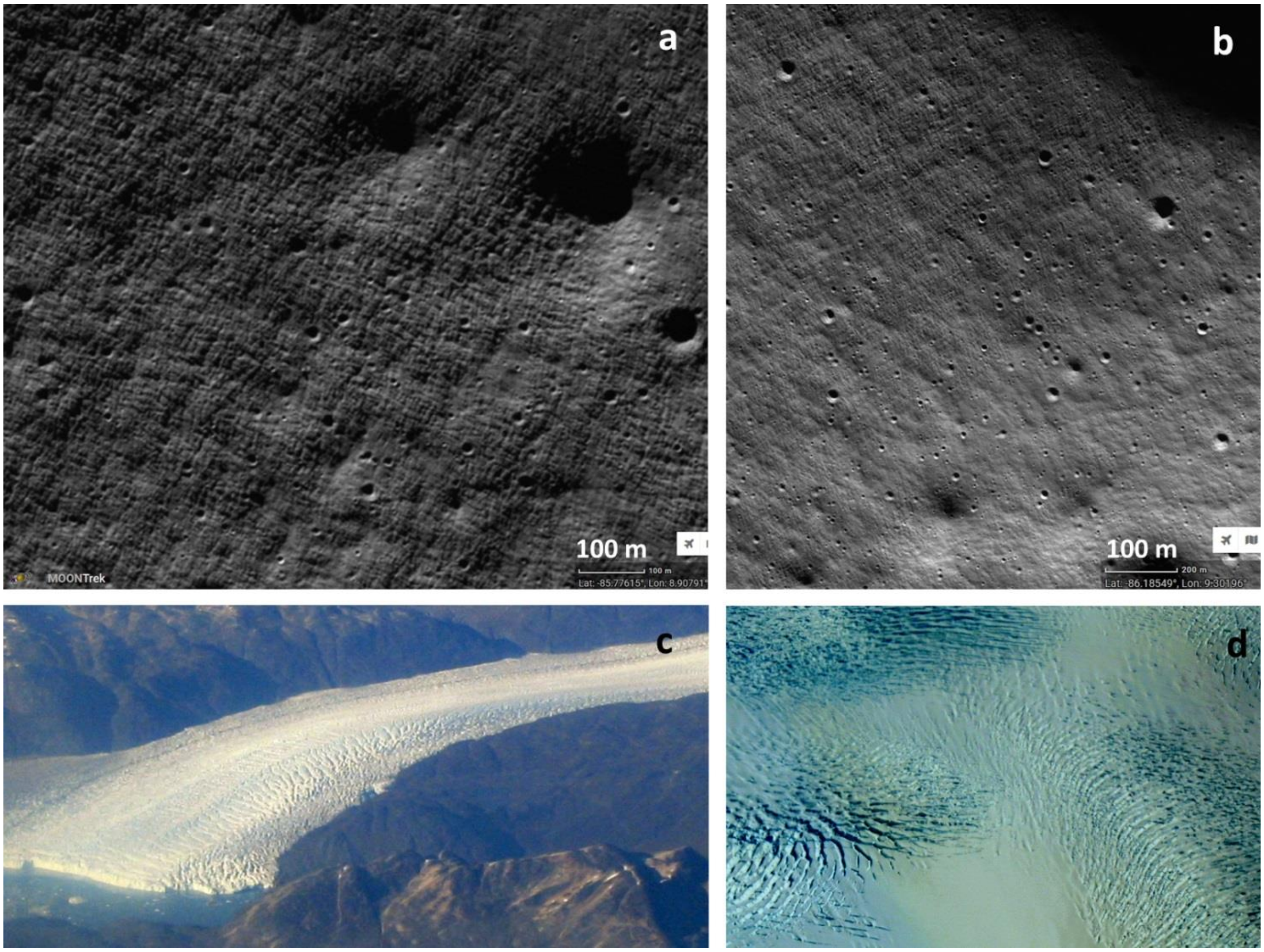

Figure 6. Enhanced fracturing in the lunar polar regions as an indicator of flowing subsurface ice. (a) Orthogonal fractures near the South Pole at coordinates 85.8°S, 8.9°E; (b) Thin, elongated fractures at 86.2°S, 9.3°E; (c) A glacier in southeastern Greenland covered with crevasses, aerial photograph, September 19, 2005; (d) Crevasses on the Greenland ice sheet (contrast enhanced), aerial photograph, May 27, 2006. (a), (b) – NASA (MoonTrek); (c), (d) – photographs by Nick Gorkavyi.

## 4. Geomorphological Signs of Possible Rapid Moonflows

The most evident manifestation of polar volatile abundance may be traces of flows caused by impact-induced melting of subsurface ice. However, liquid in a vacuum boils and rapidly evaporates or freezes. Therefore, such a flow can exist only briefly, and it would not be a river of water but rather something resembling a mudflow or mudslide.

Schmitt (2026) has found that impact shock resets the maturity index (Is/FeO, Morris, 1976) of regolith. Such shock pressures may be sufficient to locally and temporarily melt ice particles or ice matrices in polar regolith. We suggest that mixtures of regolith and liquid water may become mobile. We refer to this phenomenon as “moonflow”. The liquid water within such a mixture would boil, and along with released solar wind volatiles, would reduce the viscosity of the host regolith and potentially create moonflows. Craters that generate narrow flows forming channel-like features may represent a subset of rampart craters, which produce fluidized ejecta that spreads around the crater (for example, Mars—single layered ejecta crater 11 km across, 23.6 N, 101.7 E; the Moon—Tsiolkovskiy lobate form, 20.25°S, 128.94°E, crater 185 km across - Blanche et al., 2025).

If a crater is located on a horizontal surface, the moonflow would remain confined within it. After the liquid water evaporates and refreezes, a smooth regolith terrain may form on the crater floor. The surrounding icy regolith may also partially melt and sublimate, producing unusual surface irregularities around the crater.

If a crater forms on a slope, a short-lived moonflow may occur, leaving behind a long-lived channel or groove on the hillside. Near the lunar North Pole, we have identified several channels that appear to be traces of past liquid flows originating from small craters (Figures 7 and 8). The channel shown in Figure 7 was apparently produced by the crater up to three hundred meters in size in the upper part of the image.

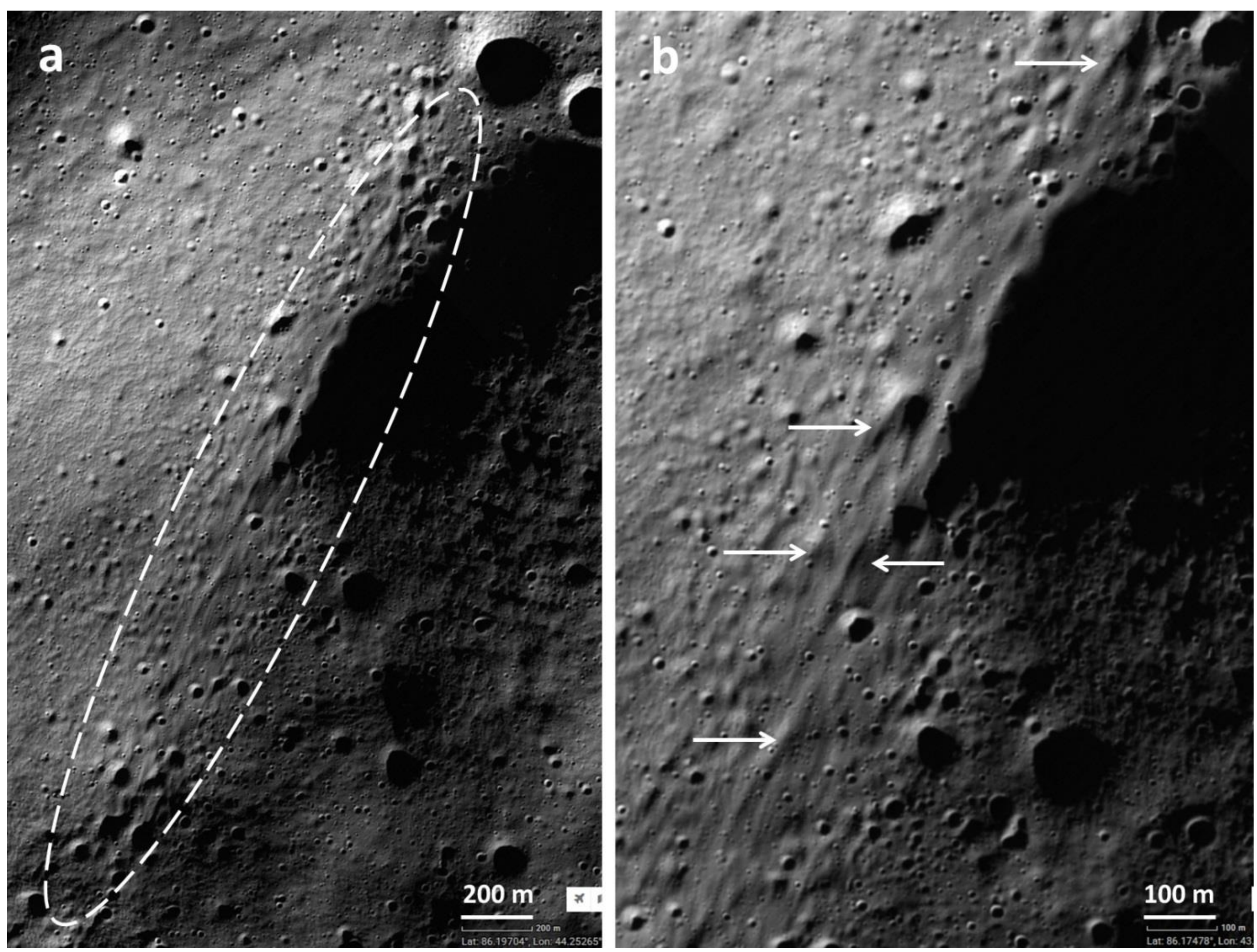

Figure 7. An ancient channel near the lunar North Pole (86.19°N, 43.8°E), possibly associated with a short-lived moonflow caused by impact melting of ice and release of volatiles. (a) The channel is outlined in white; (b) The central part of the channel. Arrows indicate V-shaped structures likely formed by rapid flow around obstacles. NASA (MoonTrek).

This flow cannot be volcanic lava or impact-melted rock because:

a. The flow is located in the mountainous region of the lunar North Pole, where there is as yet identified volcanic activity or mare-style lava flooding.

b. Craters with diameters of only a few hundreds of meters are too small to generate lava flows of sufficient volume to produce the observed features. It cannot be due to impact-melt as such flows would then be observed near many polar craters.
c. There is no depositional apron at the end of the flow, typical of impact-melt flows observed near large craters several kilometers in diameter (e.g., near Mandel'shtam F, a 15-km crater in the farside highlands, or Petavius B, a 33-km crater).

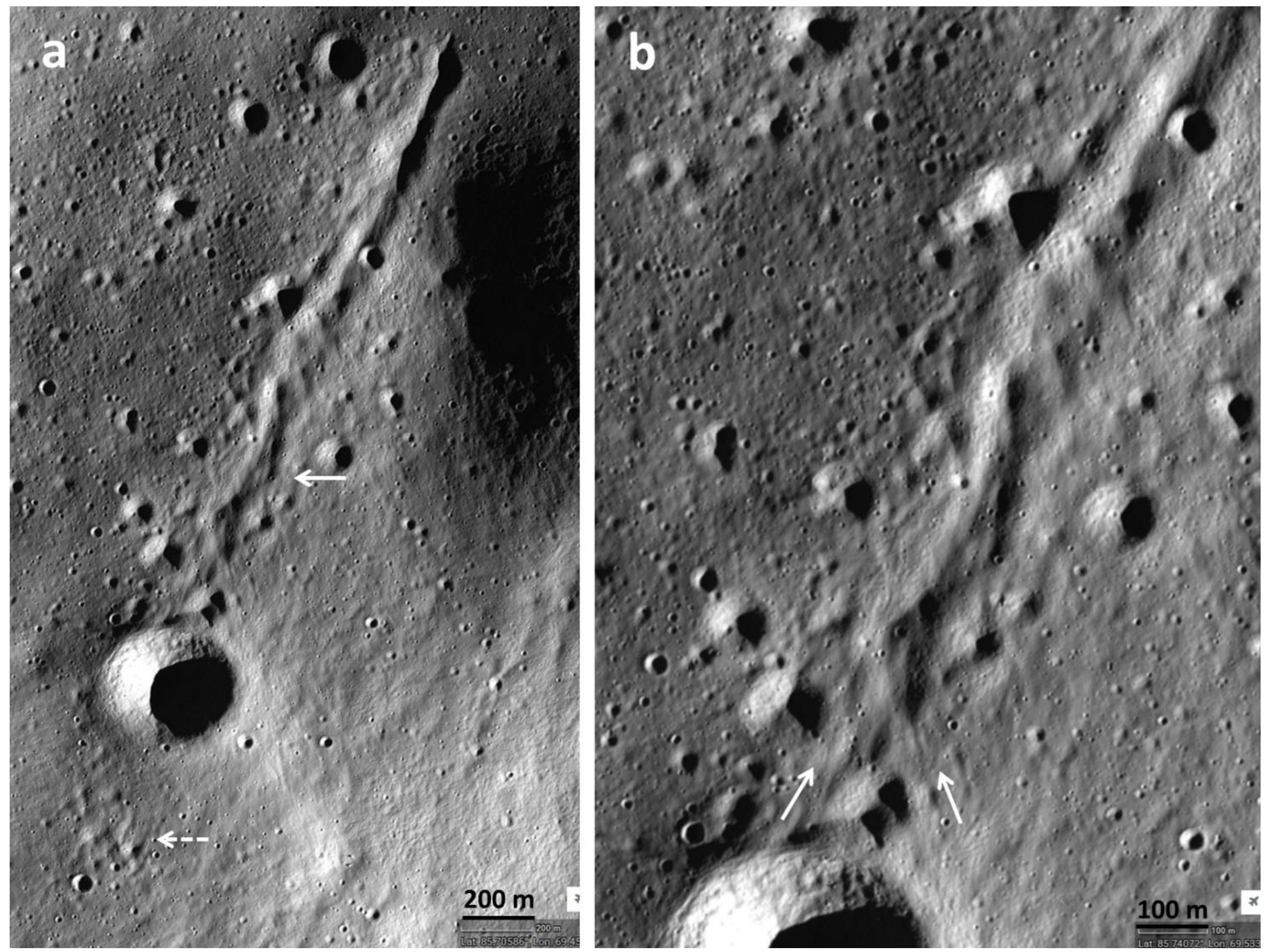

Figure 8. An ancient channel near the lunar North Pole close to a 300-m crater (85.7°N, 69.45°E), possibly associated with impact-induced melting of ice and volatiles. (a) Solid arrow indicates the moonflow channel. Dashed arrow marks terrain possibly related to partial icy regolith melting and uneven ground subsidence. (b) Close-up view of the channel near the crater. Arrows indicate downslope flow directions from the crater walls. NASA (MoonTrek).

Let us estimate the velocity of a moonflow. For an elevation drop of $H$ =100 meters, the flow velocity could reach approximately $V = \sqrt{2gH}$ ≈18 m/s (≈65 km/h), neglecting viscosity and friction. At such a speed, within two minutes the flow could carve a channel more than two kilometers long, comparable to the grooves in Figure 7 and Figure 8.

Accounting for friction and viscosity would reduce the velocity and increase the transit time, likely to several minutes, although release of other volatiles (H, N, He) may compensate for these factors. Along with the insulating nature of incorporated regolith, we assume that the amount of melted water resulting from the collision would be sufficiently large to prevent full evaporation within a few minutes, so as to enable moonflow mobility. If the channel width is about 100 meters, the flowing layer must be at least several meters thick. This implies that evaporation would begin at the flow surface, while inside the flow — where pressure as well as insulation is non-zero due to the weight of the mass and lunar gravity — evaporation would be limited. The pressure produced by regolith with density $\rho = 2$ g/cm³ at a depth h = 10 m is:

$$P = g\rho h = 0.32\ atm \qquad (1)$$

where $g = 1.62\ m/s^2$.

A more realistic estimate of the flow velocity, accounting for friction along a rough surface, can be obtained from the expression given by Kleinhans (2005):

$$V = \sqrt{\frac{8ghH}{fL}} = \frac{2.3}{\sqrt{f}}\ m/s = 4 - 7\ m/s \qquad (2)$$

where g=1.62 m/s$^2$; h is the flow depth (assumed to be ∼10 m), H is the elevation drop (assumed to be ~100 m), and L is the channel length. For the estimate in Eq. (2), we assume H/L=1/25 and a friction coefficient f=0.1− 0.3**.** The resulting flow velocity is comparable to the speed of a running person. Assuming channel lengths of approximately 2.5 km for the channel shown in Figure 7 and 1.7 km for that in Figure 8, the corresponding formation times are estimated to be about 6–10 minutes and 4–7 minutes, respectively. Interestingly, the flow velocity given by Eq. (2) is close to the propagation speed of shallow-water gravity waves, $\sqrt{gh}$. Consequently, the Froude number is expected to be close to unity. Under such conditions, a flow encountering an obstacle tends to split into two oblique waves, forming a characteristic V-shaped pattern. Similar wave structures are commonly observed when shallow streams flowing over pavement encounter small stones, or in the wakes of boats moving through water. In this context, we note the V-shaped features indicated by arrows in Figure 8. They suggest that the flow originated from the large crater.

These hypothetical considerations require confirmation through detailed modelling of boiling multiphase moonflow dynamics and dedicated experimental studies.

Analysis of the North and South Poles indicates that such channels are rarer and more degraded at the South Pole (Figure 9 a,b). Notably, Mars also exhibits numerous outflow channels that provide evidence of icy regolith melting triggered by asteroid impacts (Figure 9). The innovation of this article lies in suggesting that these phenomena could also occur on the Moon under special circumstances, in spite of the extremely low surface pressure on the order of 0.3 nPa.

The rarity of moonflow channels can be explained by the need for several simultaneous conditions:

• Shallow and abundant regolith ice (either as particles of matrix), typically possible only in polar regions.

• A crater large enough to penetrate to icy regolith, but not so large that it traps all released water within its cavity.

• The crater must be located on a slope, allowing gravity to initiate flow and channel formation.

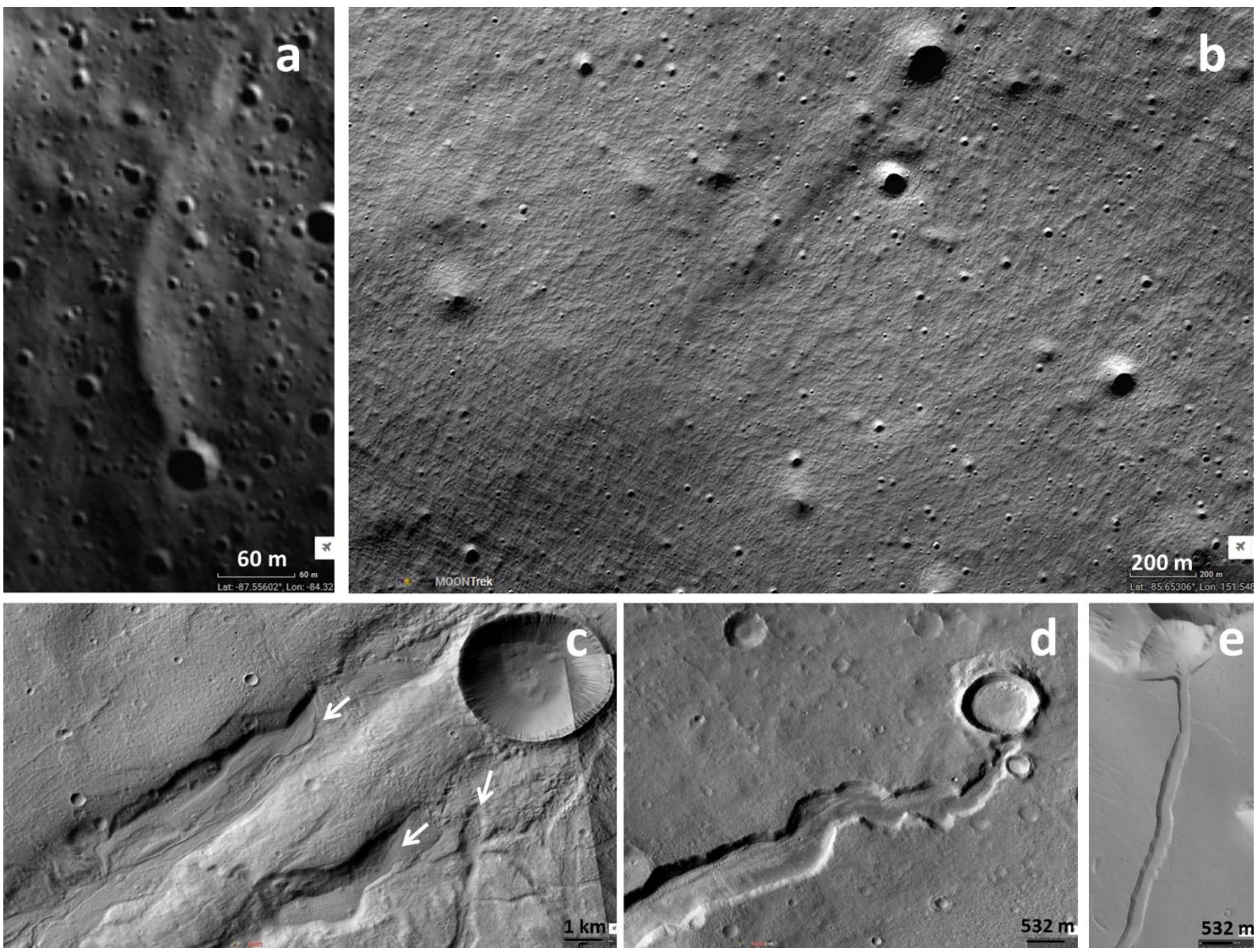

Figure 9. Degraded flow channels near the lunar South Pole and on Mars. (a) Near a lunar crater at 87.556°S, 84.32°W; (b) Near a lunar crater at 85.65°S, 151.55°E; (c) A Martian rampart crater (32.13°S, 95.0°E), 4 km in diameter, with channels (arrows) likely formed by fluidized flows or creeping glaciers; (d) Water outflow from a ~1-km Martian crater (32.706°N, 18.329°E), forming a narrow channel with steep walls that widens downstream due to icy regolith degradation. The channel walls are approximately 100–200 meters high, possibly indicating icy regolith depth; (e) A Martian landslide (18.58°N, 126.32°W) that may have triggered subsurface icy regolith melting and generated a long channel. NASA (MoonTrek; MarsTrek).

## 5. Dry avalanches in the Taurus–Littrow Valley

The Apollo 17 expedition in December 1972 explored the Taurus–Littrow Valley, part of which is shown in Figure 10 (Schmitt, 1973, Schmitt et al., 2017). On the left, the linear structure of the Lee–Lincoln Scarp is highlighted; on the right is a debris flow deposit characterized by a relatively young, smooth surface.

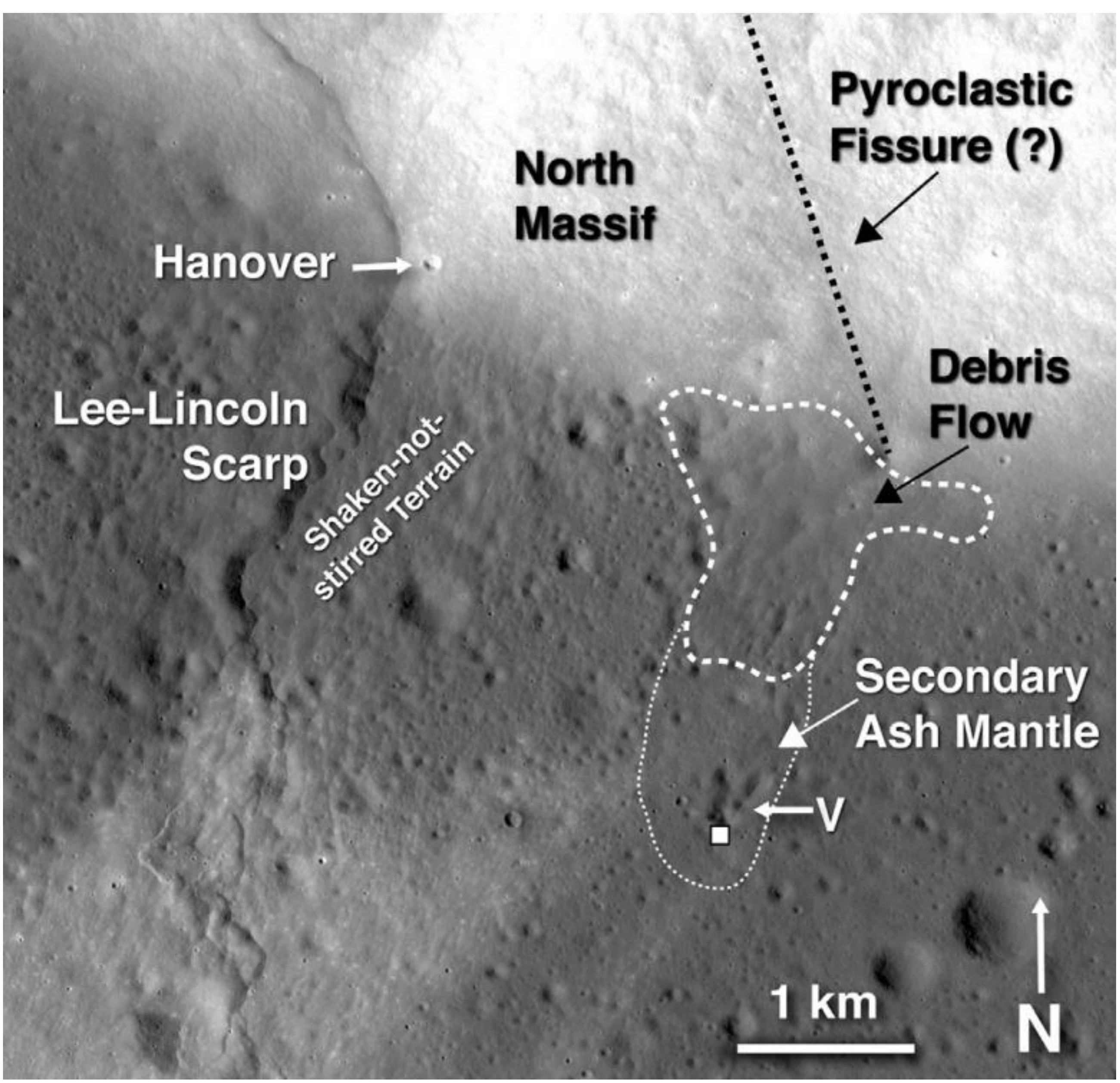

Figure 10. This image (Schmitt et al., 2017), shows the Lee–Lincoln Scarp near the base of the North Massif. The upper side of the fault (the hanging wall) appears smooth, while the lower side (the footwall) is much rougher. The view spans about 10 km from left to right. Several important features are highlighted. A line of reduced radar reflections on the southern slope of the North Massif (shown as a black dotted line) likely marks a volcanic fissure. At the southern end of the fissure, there is a distinctive cone-shaped depression without a raised rim, likely formed by subsidence. A possible debris flow deposit is outlined with a white dotted line, and a thinner dotted line shows a secondary ash layer. This ash layer was sampled at LRV Station 7 (marked by a white square), where sample 75115 contained about 16% orange glass mixed with black ash. The letter "V" marks the location of Victory Crater. LROC Image M104311715LR.

In the works of Schmitt et al. (2017) and Schmitt (2026), the light mantle deposits of the Taurus–Littrow Valley were studied in detail and interpreted as products of regolith mass wasting (fluidized avalanche) from the slopes of surrounding massifs. A key feature of such processes is the absence of stable flow channelization. The motion of dry regolith occurs in a granular medium regime, where weak interparticle interactions, probably fluidization from released solar wind volatiles, and low gravity lead to rapid distal spreading of the flow and the formation of characteristic distal plume-shaped structures.

The data from Taurus–Littrow show that even with substantial material volumes and the presence of inclined surfaces, dry avalanches do not form stable channels. On the contrary, they spread directly away from their source, forming linear blanket-like deposits. Even in cases of long-runout landslides, the morphology of the deposits does not include the formation of narrow linear channels (Figures 11 and 12) but rather show a surface ridge and trough texture.

Schmitt et al. (2017), integrating astronaut field observations, photographs, LROC NAC data, M³ Chandrayaan-1 spectra, and Mini-RF radar data, showed that the younger of the two deposits in Figure 11 likely formed as a result of a seismic trigger — a displacement along the Lee–Lincoln Scarp between 27 and 32 million years ago (Schmitt, 2026). The morphology of these avalanches is characteristic of fluidized dry granular flows under lunar gravity and vacuum conditions: broad blanket deposits spreading directly across the valley floor and perpendicular to the parent slope as a relatively thin layer (meters to tens of meters) with high albedo due to low maturity. They do not form deeply incised, narrow, and elongated channels. The transport mechanism is likely volatile a "fluidization" of dry granular material without the involvement of a liquid phase. Similar dry avalanches are also observed off the walls of large lunar craters, where they likewise appear as broad aprons or finger-like deposits.

The narrow and elongated channels observed near the lunar polar regions (Figures 7–9) are characterized by a high degree of linearity and directional consistency (similar to the Taurus-Littrow light mantles), limited width over significant lengths, and the absence of well-developed fan-shaped terminal deposits. Although the channels in Figures 7 and 8 and the avalanche traces in Figures 10–12 differ noticeably, we cannot claim that the channels in Figures 7 and 8 could not have been formed by fluidized dry landslides or downslope slumping. It should be noted that avalanches in polar regions may be more fluid than those near the equator due to the higher abundance of volatiles near the poles.

It should be emphasized that the data presented in Chapters 4 and 5 here do not constitute definitive evidence for a water-related origin of the polar channels. However, there is sufficient information that supports the need of additional research on this topic, within the context of future lunar exploration. An interesting question, relevant for both modelling and in-situ studies, arises about the influence of water (ice) content on the occurrence of lunar avalanches. Assuming that the idea of ice triggering moonflows is supported by more evidence, what is the ice fraction threshold that would make dry fluidized flows transition into liquid mudflows?

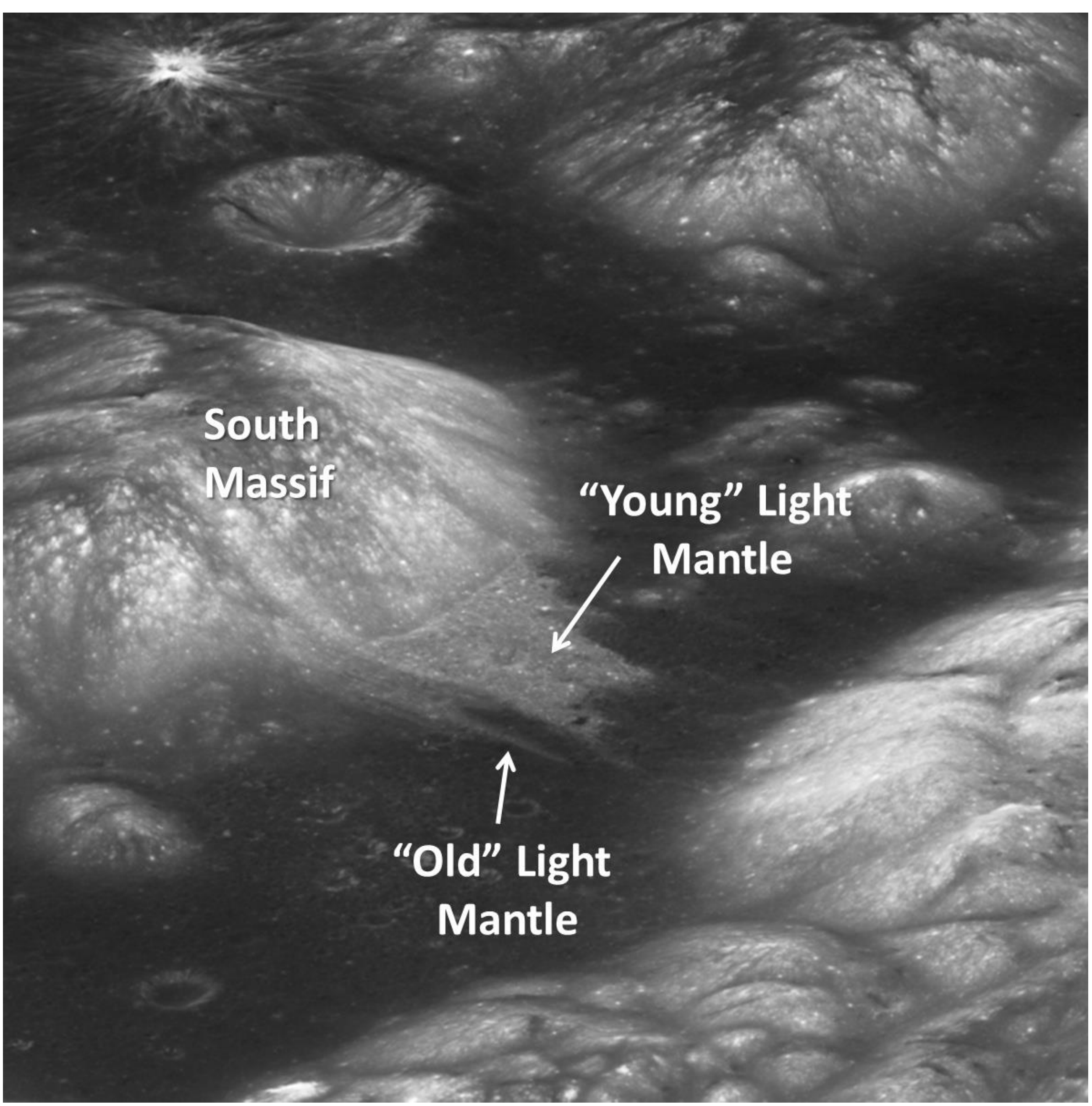

Figure 11. Another image (from Schmitt et al., 2017 with modifications), taken under high Sun from east to west, shows Taurus–Littrow Valley. Near the base of the South Massif (center-left), a plume-like avalanche deposit can be seen. The image also suggests that a darker, older surface layer lies beneath a brighter, younger one. LROC Image M1182232465.

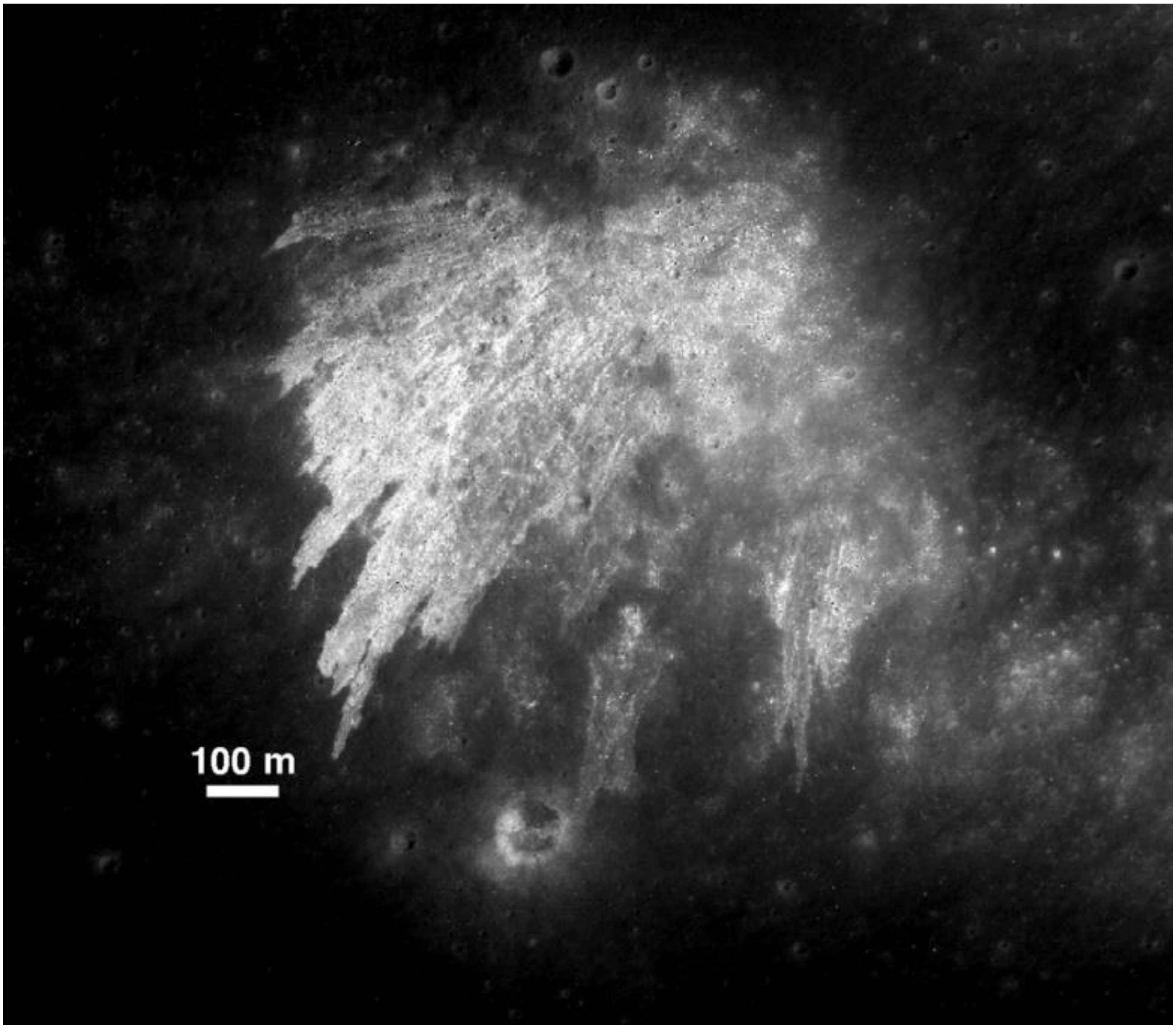

Figure 12. The southeastern peak of the Sculptured Hills shows a “paint-splatter” pattern created by granular debris flows of different ages, distinguished by their brightness. Thin bright lines within these flows are rampart-like ridges of boulders formed along their edges. (NASA, from Schmitt et al., 2017).

## 6. Thermodynamics of Lunar Ice Melting

Let us consider the overall energy balance of an asteroid impact. As examples, we will consider a 250-meter crater (rim diameter), which likely caused the formation of the channel shown in Figure 7, and a 350-meter crater corresponding to the channel in Figure 8. To calculate the impact parameters, we will use the formulas from Holsapple (2003), specifically his full equation (7), which describes craters in both the gravity and strength regimes, and equation (9), which estimates the volume of material melted during an asteroid impact. The calculation results for the two crater sizes are presented in Table 5. According to Ivanov (2001), the average impact velocity of asteroids on the Moon is 16 km/s, but the velocity distribution is broad. Therefore, we also considered velocities of 8 km/s and 32 km/s as quite plausible options.

Table 5. Asteroid Impact Parameters and Energy for 250-m and 350-m Lunar Craters

| Size of crater | V_aster (km/s) | R_aster (m) | M_aster (tn) | Energy of asteroid (erg) | Crater volume ($m^3$) | Excavation mass (tn) | Specific energy (erg/g) | Melted volume ($m^3$) |
|---|---|---|---|---|---|---|---|---|
| 250 m | 8 | 7.2 | 2810 | $9.0*10^{20}$ | $6.8*10^5$ | $1.0*10^6$ | $9.0*10^8$ | $2.0*10^3$ |
| 250 m | 16 | 5.5 | 1250 | $1.6*10^{21}$ | $6.8*10^5$ | $1.0*10^6$ | $1.6*10^9$ | $9.9*10^3$ |
| 250 m | 32 | 4.1 | 520 | $2.7*10^{21}$ | $6.8*10^5$ | $1.0*10^6$ | $2.7*10^9$ | $1.8*10^4$ |
| 350 m | 8 | 10.9 | 9760 | $3.1*10^{21}$ | $1.9*10^6$ | $2.8*10^6$ | $1.1*10^9$ | $6.9*10^3$ |
| 350 m | 16 | 8.2 | 4160 | $5.3*10^{21}$ | $1.9*10^6$ | $2.8*10^6$ | $1.9*10^9$ | $3.3*10^4$ |
| 350 m | 32 | 6.2 | 1800 | $9.2*10^{21}$ | $1.9*10^6$ | $2.8*10^6$ | $3.2*10^9$ | $5.7*10^4$ |

It should be noted that since the crater diameters are fixed, the crater volumes and excavation mass remain the same for different velocities, because the crater depth formula (1/4.8 of the rim diameter) does not depend on impact velocity, according to the Holsapple (2003) model. This relationship can also be found in Melosh (1989).The following parameters were used in the calculations (Holsapple, 2003): g=1.62 m/s², density of the impacting C-type asteroid 1.8 g/cm³. For the lunar regolith, a density of ρ = 1.5 g/cm³ and a strength of $10^5$ dynes/cm² were adopted. When determining impact crater shapes, it was assumed that the crater formed in “dry soils (some cohesion)” or “soft rock” (Holsapple, 2003).

Thus, on average, each gram of affected regolith and basalt would receive a specific impact kinetic energy of: $9\times10^8$ - $3.2\times10^9$ erg/g in addition to subsequently release potential energy (Schmitt, 2026). Impact energy is spent on mechanical effects (fracturing, compression, ejecta excavation) and thermal effects (heating and melting of the regolith and basalt). Experiments and simulations indicate that no more than half of the impact energy goes into heating and melting (Lindsay, 1976). How does subsurface ice modify the mechanical and thermal effects? The mechanical strength of ice is comparable to that of rock, so its presence does not dramatically alter fragmentation behaviour (however, vapor clouds generated during impact may enhance ejecta dispersal). The thermal balance changes significantly, since melting and vaporization temperatures of ice and water differ strongly from the melting temperature of regolith grains.

To determine the amount of melted silicates, the melt energy for silicates of $5\times10^{10}$ erg/g was used (Holsapple, 2003). This gives a volume of melted material equal to 1.3 times the asteroid volume at an impact velocity of 8 km/s (see also O’Keefe & Ahrens, 1977), 14 times the asteroid volume at 16 km/s, and 57 times at 32 km/s.

How much material/water is required to form a channel 100 meters wide? Let us estimate the necessary volume of liquid mass as: $100\times100\times10$ m=$10^5$ $m^3$. Assuming that liquid water constitutes 10% of this volume, the impact would release approximately $10^4$ tons of liquid water.

The latent heat of fusion of ice is $3.3 \times 10^{10}$ erg/g**,** which is close to the value of $5 \times 10^{10}$ erg/g adopted in our calculations. Therefore, the estimated melt volume given in the last column of Table 5 may also be regarded as an estimate of the volume of melted ice. The temperature of the ice-bearing regolith is expected to remain close to the melting point until most of the ice has melted.

Table 5 shows that for impact velocities of 8 km/s and higher, the amount of liquid water released would range from approximately 2,000 to 50,000 tons**.** Even for an impact velocity of 8 km/s**,** the estimated amount of water is 2,000–7,000 tons**,** which may still be sufficient to trigger a debris avalanche. For an impact velocity of 16 km/s, the released water represents only 1.5–1.7% of the total crater volume.

These calculations indicate that the impact of a small asteroid capable of producing a crater approximately 300 m in diameter could generate a moonflow consisting of regolith mixed with meltwater.

An interesting case is the Martian channel caused by liquid water (or a liquid water–sand mixture) shown in Figure 7e. That flow was not triggered by an asteroid impact but by a landslide descending several hundred meters. The specific gravitational energy per gram of soil in that case is: $gh \sim 10^6$–$10^7$ erg/g — far smaller than the energy involved in lunar asteroid impacts. Nevertheless, this appears sufficient to melt ice along fracture surfaces and sliding planes within the landslide mass.

## 7. Conclusion

This work suggests that the existence of subsurface icy regolith and the abundance of volatiles in mountainous polar regions is not only important for life support and rocket fuel production, but also because it introduces geotechnical aspects whose associated risks need to be assessed and mitigated for the implementation of the Artemis program.

Possible operational implications of Artemis activities on the Moon associated with subsurface icy regolith may include:

- Landings and liftoffs of powerful rockets (short-term, repeated disturbances);
- Prolonged thermal influence of a habitable base;
- Prolonged thermal influence of operating a nuclear reactor.
- Source of water
- Source of cooling
- Source of road and path macadam

These factors may significantly affect the stability of volatile-rich layers, increasing the risk of subsidence and fracturing typical of icy regolith thaw and glacier motion. In addition, asteroid impact hazards near a base may be amplified, since such impacts could trigger rapid moonflows if activities and facilities are located near significant slopes, expanding the danger zone around the impact site.

The hypothesis of a substantial subsurface icy regolith layer in mountainous polar regions, discussed in this paper, is supported by numerous observations, experimental findings, and theoretical calculations. While these results are not conclusive, they support additional investigations in the form of remote sensing, geophysical prospecting and geotechnical assessments, including simulant evaluations as a function of ice content and differences between particle and matrix ice.

In particular, combined geophysical campaigns involving gravimetry, radar (GPR) and seismics, magnetic and electro-properties methodologies, together with drilling campaigns with core sampling to depths of several tens of meters may allow a detailed mapping of icy regolith-rich areas around the South Pole of the Moon and elsewhere, which would reduce risks associated with the construction and maintenance of a Lunar base.

## Acknowledgments

This research was supported by the Science Systems and Applications Incorporated (SSAI) Internal Research & Development (IRAD) program.

## References

Akhmanova, M.V., Dement'ev, B.V., Markov, M.N., 1979. Possible water in Luna 24 regolith from the Sea of Crises. Geochemistry International 15, 166–168.

Alfvén, H., Arrhenius, G. Origin and evolution of the earth-moon system. The Moon, 5, 210–230 (1972). https://doi.org/10.1007/BF00562115

Arnold, J.R.,1979. Ice in the lunar polar regions. J. Geophys. Res. 84, 5659-5668.

Blance, A.J., Lennox, A.R., Rothery, D.A., Balme, M., Wright, J., Galluzzi, V., Conway, S.J., 2025. Lobate forms around craters on the Moon and Mercury: Origin from landslides, ejecta flows and modification stage collapse. Journal of Geophysical Research: Planets 130, e2025JE008980. https://doi.org/10.1029/2025JE008980

Cadenhead, D.A., Buergel, W.G., 1974a. Gas interaction studies with lunar orange soil 74220,29. In: Abstracts of the 5th Lunar and Planetary Science Conference, p. 100.

Cadenhead, D.A., Buergel, W.G., 1974b. The interaction of hydrogen with Taurus–Littrow orange soil. In: Proceedings of the 5th Lunar Conference. Geochimica et Cosmochimica Acta, Suppl. 3, 2287–2300.

Chauhan, P., Chauhan, M., Verma, P.A., Sharma, S., Bhattacharya, S., Dagar, A., Amitabh, Patil, A.N., Parashar, A.K., Kumar, A., Desai, N., Karidhal, R., Kumar, A., 2021. Unambiguous detection of OH and $H_2O$ on the Moon from Chandrayaan-2 Imaging Infrared Spectrometer reflectance data using the 3 μm hydration feature. Current Science 121, 391–401. https://doi.org/10.18520/CS/V121/I3/391-401.

Colaprete, A., et al., 2010. Detection of water in the LCROSS ejecta plume. Science 330, 463–468. https://doi.org/10.1126/science.1186986

Crotts, A., 2011. Water on the Moon I: Historical overview. Astronomical Review 6 (8), 4–20. https://doi.org/10.48550/arXiv.1205.5597

Crotts, A., 2012a. Water on the Moon II: Origins and resources. Astronomical Review 7 (1), 36–47. https://doi.org/10.48550/arXiv.1205.5598.

Crotts, A., 2012b. Water on the Moon III: Volatiles and activity. Astronomical Review 7, 53–94. https://doi.org/10.48550/arXiv.1205.5599

Fassett, C.I., Beyer, R.A., Deutsch, A.N., Hirabayashi, M., Leight, C.J., Mahanti, P., et al., 2022. Topographic diffusion revisited: Small crater lifetime on the Moon and implications for volatile exploration. Journal of Geophysical Research: Planets 127, e2022JE007510. https://doi.org/10.1029/2022JE007510.

Feldman, W.C., Maurice, S., Lawrence, D.J., Little, R.C., Lawson, S.L., Gasnault, O., Wiens, R.C., Barraclough, B.L., Elphic, R.C., Prettyman, T.H., Steinberg, J.T., Binder, A.B., 2001. Evidence for water ice near the lunar poles. Journal of Geophysical Research: Planets 106, 23231–23251. https://doi.org/10.1029/2000JE001444.

Gläser, P., Sanin, A., Williams, J.-P., Mitrofanov, I., & Oberst, J. (2021). Temperatures near the lunar poles and their correlation with hydrogen predicted by LEND. Journal of Geophysical Research: Planets, 126, e2020JE006598. https://doi.org/10.1029/2020JE006598

Gorkavyi, N., 2022. Investigation of geomorphological signatures of permafrost in the polar lunar areas with VIPER. NASA Technical Report. https://ntrs.nasa.gov/citations/20220018310

Gorkavyi, N., 2023. Origin of the Moon and lunar water. Earth and Planetary Science 2 (2), 85–98. https://doi.org/10.36956/eps.v2i2.940

Hartmann, W. K., Davis, D. R.,  Satellite-sized planetesimals and lunar origin, Icarus, 24, 4, (1975), 504-515, ISSN 0019-1035, https://doi.org/10.1016/0019-1035(75)90070-6.

Hauri, E. H., T. Weinreich, A. E. Saal, M. C. Rutherford, and J. A. Van (2011), High Pre-Eruptive Water Contents Preserved in Lunar Melt Inclusions. *Science,* **333**, 213-215.

Holsapple, K.A. Users Manual: Crater Sizes from Explosions or Impacts. 2003. https://www.lpi.usra.edu/lunar/tools/lunarcratercalc/theory.pdf

Honniball, C.I., Lucey, P.G., Li, S. *et al.* (2021a) Molecular water detected on the sunlit Moon by SOFIA. *Nature, Astron.* **5**, 121–127. https://doi.org/10.1038/s41550-020-01222-x

Honniball, C.I., Lucey, P.G., Petro, N.E., Li, S., Young, K.E., (2021b). Enhanced hydration at craters with central peaks detected by ground-based observations. In: 52nd Lunar and Planetary Science Conference, LPI Contribution No. 2548, Abstract 1256.

Ivanov, B.A. Mars/Moon cratering rate ration estimates. Space Science Reviews. 96, 87–104 (2001). https://doi.org/10.1023/A:1011941121102

Kletetschka, G., Klokočník, J., Hasson, N., et al., 2022. Distribution of water phase near the poles of the Moon from gravity aspects. Scientific Reports 12, 4501. https://doi.org/10.1038/s41598-022-08305-x

Killen, R. M., A. E.Potter, D. M.Hurley, C.Plymate, and S.Naidu (2010), Observations of the lunar impact plume from the LCROSS event, Geophys. Res. Lett., 37, L23201, doi:10.1029/2010GL045508.

Kleinhans, M. G. (2005), Flow discharge and sediment transport models for estimating a minimum timescale of hydrological activity and channel and delta formation on Mars, J. Geophys. Res., 110, E12003

Klima, R., Cahill, J., Hagerty, J., et al., 2013. Remote detection of magmatic water in Bullialdus crater on the Moon. Nature Geoscience 6, 737–741. https://doi.org/10.1038/ngeo1909

Kling, A.M., Greer, J., Thompson, M.S., Heck, P.R., Isheim, D., Seidman, D.N., 2025. Nanoscale reservoirs store solar wind-derived water on the lunar surface. Earth and Planetary Science Letters 651, 119178. https://doi.org/10.1016/j.epsl.2024.119178

Lawrence, D.J., Feldman, W.C., Elphic, R.C., Hagerty, J.J., Maurice, S., McKinney, G.W., Prettyman, T.H., 2006. Improved modeling of Lunar Prospector neutron spectrometer data: Implications for hydrogen deposits at the lunar poles. Journal of Geophysical Research: Planets 111, E08001. https://doi.org/10.1029/2005JE002637

Lawrence, D.J., Peplowski, P.N., Wilson, J.T., Elphic, R.C., 2022. Global hydrogen abundances on the lunar surface. Journal of Geophysical Research: Planets 127, e2022JE007197. https://doi.org/10.1029/2022JE007197

Li, S., Milliken, R.E., 2017. Water on the surface of the Moon as seen by the Moon Mineralogy Mapper: Distribution, abundance, and origins. Science Advances 3, e1701471. https://doi.org/10.1126/sciadv.1701471

Lindsay, J. 1976. Lunar Stratigraphy and Sedimentology. Elsevier.

Lock, S. J., Stewart, S. T., Petaev, M. I., Leinhardt, Z., Mace, M. T., Jacobsen, S. B., & Ćuk, M. (2018). The origin of the Moon within a terrestrial synestia. Journal of Geophysical Research: Planets, 123, 910–951. https://doi.org/10.1002/2017JE005333

Lucey, P. G. , Ryan A. Zeigler, Lingzhi Sun, Abigail Flom, Andrea B. Mosie, Juliane Gross, Marley A. Chertok, Chiara Ferrari-Wong, Schelin M. Ireland, the ANGSA Team (2024) Infrared Spectroscopy of Lunar Core 73001: Upper Limit on Hydration in a Lunar Sample With No History of Exposure to Terrestrial Water Vapor. JGR Planets, https://doi.org/10.1029/2024JE008389

Magaña, L.O., Retherford, K.D., Byron, B.D., Hendrix, A.R., Grava, C., Mandt, K.E., Raut, U., Czajka, E., Hayne, P.O., Hurley, D.M., Gladstone, G.R., Poston, M.J., Greathouse, T.K., Pryor, W., Cahill, J.T., Stickle, A., 2022. LRO-LAMP survey of lunar south pole cold traps: Implication for the presence of condensed $H_2O$. Journal of Geophysical Research: Planets 127, e2022JE007301. https://doi.org/10.1029/2022JE007301

McConnochie, T. H., Bonnie J. Buratti, John K. Hillier, Kimberly A. Tryka, 2002. A Search for Water Ice at the Lunar Poles with Clementine Images, Icarus, Volume 156, Issue 2, Pages 335-351, ISSN 0019-1035, https://doi.org/10.1006/icar.2001.6765. (https://www.sciencedirect.com/science/article/pii/S001910350196765X)

Melosh, H.J. Impact Cratering. A Geological Process. Oxford University Press, New York; Clarendon Press, Oxford (1989)

Meyer, C. (2012) Lunar Sample Compendium, https://curator.jsc.nasa.gov/lunar/lsc/76001.pdf, for sample 76001; and other samples at https://curator.jsc.nasa.gov/lunar/lsc/index.cfm.

Moore J.M. et al. The geology of Pluto and Charon through the eyes of New Horizons. Science 351,1284-1293 (2016)

Nakajima, M., Stevenson, D. J., Investigation of the initial state of the Moon-forming disk: Bridging SPH simulations and hydrostatic models, (2014), Icarus 233, 259-267, ISSN 0019-1035, https://doi.org/10.1016/j.icarus.2014.01.008.

Nakajima, M., Stevenson, D. J., Inefficient volatile loss from the Moon-forming disk: Reconciling the giant impact hypothesis and a wet Moon, (2018). Earth and Planetary Science Letters, 487, 117-126, ISSN 0012-821X, https://doi.org/10.1016/j.epsl.2018.01.026.

Naver, E.B., Nikolajsen, K.W., Carøe, M.S., Battaglia, D., Frydenvang, J., Bizzarro, M., Jørgensen, J.S., Lefmann, K., Kaestner, A., Mannes, D.C., Cook, P., Birkedal, H., Christensen, T.E.K., Kantor, I., Poulsen, H.F., Kuhn, L.T., 2026. Direct detection of hydrogen reveals a new macroscopic crustal water reservoir on early Mars. arXiv preprint. https://doi.org/10.48550/arXiv.2601.08390

Nozette S, Lichtenberg CL, Spudis P, Bonner R, Ort W, Malaret E, Robinson M, Shoemaker EM., 1996. The Clementine bistatic radar experiment. Science ;274(5292):1495-8. doi: 10.1126/science.274.5292.1495. PMID: 8929403.

Nozette, S., P. D. Spudis, M. S. Robinson, D. B. J. Bussey, C. Lichtenberg, and R. Bonner (2001), Integration of lunar polar remote-sensing data sets: Evidence for ice at the lunar south pole, J. Geophys. Res., 106(E10), 23253–23266, doi:10.1029/2000JE001417.

Ohtake M. et al (2024) Plumes of Water Ice/Gas Mixtures Observed in the Lunar Polar Region. The Astrophysical Journal, 963, 124.

O'Keefe, J. D. & Ahrens, T. J. (1977) Impact-induced energy partitioning, melting, and vaporization on terrestrial planets. In: Lunar Science Conference, 8th, Houston, Tex., March 14-18, 1977, Proceedings. Volume 3. (A78-41551 18-91) New York, Pergamon Press, Inc., 1977, p. 3357-3374.

Osborne, N.S., Stimson, H.E. and Ginnings D.C. (1939) Measurements of heat capacity and heat of vaporization of water in the range 0° to 100°C. Part of Journal of Research of the National Bureau of Standards. 23, 197-260

Pieters, C.M., Goswami, J.N., Clark, R.N., Annadurai, M., Boardman, J., Buratti, B., Combe, J.P., Dyar, M.D., Green, R., Head, J.W., Hibbitts, C., Hicks, M., Isaacson, P., Klima, R., Kramer, G., Kumar, S., Livo, E., Lundeen, S., Malaret, E., McCord, T., Mustard, J., Nettles, J., Petro, N., Runyon, C., Staid, M., Sunshine, J., Taylor, L.A., Tompkins, S., Varanasi, P., 2009. Character and spatial distribution of OH/$H_2O$ on the surface of the Moon seen by $M^3$ on Chandrayaan-1. Science 326, 568–572. https://doi.org/10.1126/science.1178658

Ruskol, E.L., 1975. Origin of the Moon. Nauka, Moscow.

Safronov, V.S.: Evolution of the protoplanetary cloud and formation of the Earth and the planets. Washington, DC, 1972, NASA.

Schmitt, H.H., 1973. Apollo 17 report on the Valley of Taurus-Littrow: A geological investigation of the valley visited on the last Apollo mission to the Moon. Science 182, 681–690. https://doi.org/10.1126/science.182.4113.681

Schmitt, H.H., 1990. Apollo 17 orange soil: Interpretation of geologic setting. In: Delano, J.W., Heiken, G.H. (Eds.), Workshop on Lunar Volcanic Glasses: Scientific and Resource Potential. LPI Technical Report 90-02, Lunar and Planetary Institute, Houston, TX, p. 56.

Schmitt, H. H. (2016) Symplectites in dunite 72425 and troctolite 76535 indicate mantle overturn beneath lunar near side (2016) *LPSC XLVII*, abstract 2339.

Schmitt, H. H. (2017) Geology of Shorty Crater pyroclastic ash deposits: 44 years and counting for Apollo 17. LPSC XLVIII, abstract 2017.

Schmitt, H.H., Petro, N.E., Wells, R.A., Robinson, M.S., Weiss, B.P., Mercer, C.M., 2017. Revisiting the field geology of Taurus–Littrow. Icarus 298, 2–33. https://doi.org/10.1016/j.icarus.2016.11.042

Schmitt, H.H., 2026, HISTORY IN THE DUST. Regolith of Taurus–Littrow: Ejecta Zones, Source Craters, Ages and Implications. https://www.americasuncommonsense.com/1-apollo-17-diary-of-the-12th-man/c-chapters-10-18/a-section-1/#Sect7.0

Shearer, C. K., F. M. McCubben, S. Eckley, S. B. Simon, A. Meshik, F. McDonald, H. H. Schmitt, and the ANGSA Team (2024) Apollo Next Generation Sample Analysis (ANGSA): An Apollo participating scientist program to prepare the lunar sample community for Artemis. Space Sci. Rev. 220, 62. https://doi.org/10.1007/s11214-024-01094-x.

Sinha, R.K., Bharti, R.R., Acharyya, K. et al. Subsurface ice in doubly shadowed craters as revealed by Chandrayaan-2 dual frequency synthetic aperture radar. npj Space Explor. 2, 22 (2026). https://doi.org/10.1038/s44453-026-00038-9

Sori, M.M., Byrne, Sh., Hamilton, Ch. W., Landis, M.E. (2016) Viscous flow rates of icy topography on the north polar layered deposits of Mars. GRL, 43, 541-549.

Speyerer, E. J.; R. Z. Povilaitis, M. S. Robinson, P. C. Thomas, R. V. Wagner, (2016). Quantifying crater production and regolith overturn on the Moon with temporal imaging. *Nature*. **538** (7624): 215–218

Spudis,P., Nozette, S., Lichtenberg, C., Bonner, R., Ort, W., Malaret, E., Robinson, M., Shoemaker, E. 1998. The clementine bistatic radar experiment: Evidence for ice on the moon. Solar System Research, 32, 17-22

Strycker, P., Chanover, N., Miller, C. *et al.* Characterization of the LCROSS impact plume from a ground-based imaging detection. Nature Communications. 4, 2620 (2013). https://doi.org/10.1038/ncomms3620

Toyokawa, K., Haruyama, J., Iwata T., Nozawa, H. (2024) Water ice particles detected by SELENE's Spectral Profiler at lunar shadowed regions in various local times and latitudes. Earth and Planetary Science Letters, 648, 119065.

Watson, K., H. Brown and B. C. Murray (1961) The behavior of volatiles on the lunar surface, Jour. Geophys. Res., 66**,** 3033-3045 https://doi.org/10.1029/JZ066i009p03033.

Wilhelms, D. E. (1987) The Geologici History of the Moon, USGS Prof. Paper 1348, 302p.

Wood, J. A. (1986). Moon over Mauna Loa - a review of hypotheses of formation of earth's moon (pp. p. 17–55.). Lunar and Planetary Institute.